\documentclass[11pt,a4paper,oneside]{article}
\usepackage[utf8]{inputenc}
\usepackage{amsmath}
\usepackage{amsfonts}
\usepackage{amssymb}
\usepackage{indentfirst}
\usepackage{graphicx}
\usepackage[outdir=./]{epstopdf}
\graphicspath{{./}}
\usepackage{authblk}
\usepackage{multirow}
\usepackage{longtable}
\usepackage{tabularx}
\usepackage{caption,booktabs}
\usepackage{ltablex}
\usepackage{subfigure}
\usepackage{adjustbox}
\usepackage[font=normalsize]{subfig}
\DeclareMathOperator{\sign}{sign}

\title{A comparative study of implicit 
Jacobian-free Rosenbrock-Wanner, ESDIRK and BDF methods for unsteady flow simulation with high-order flux reconstruction formulations}
\author{Lai Wang\thanks{PhD candidate. Email: bx58858@umbc.edu} \ and Meilin Yu\thanks{Assistant Professor. Corresponding Author. Email: mlyu@umbc.edu}
}
\affil{Department of Mechanical Engineering, \\
University of Maryland, Baltimore County, Baltimore, MD 21250}
\date{\vspace{-10ex}}
\begin{document}
\maketitle
	\section*{Abstract}
	We conduct a comparative study of the Jacobian-free  linearly implicit Rosenbrock-Wanner (ROW) methods, the explicit first stage, singly diagonally implicit Runge-Kutta (ESDIRK) methods, and the second-order backward differentiation formula (BDF2) for unsteady flow simulation using spatially high-order flux reconstruction/correction procedure via reconstruction (FR/CPR) formulations. The pseudo-transient continuation is employed to solve the nonlinear systems resulting from the temporal discretizations with ESDIRK and BDF2. A Jacobian-free implementation of the restarted generalized minimal residual method (GMRES) solver is employed with a low storage element-Jacobi preconditioner to solve linear systems, including those in linearly implicit ROW methods and those from linearization of the nonlinear systems in ESDIRK and BDF2 methods. We observe that all ROW and ESDIRK schemes (from second order to fourth order) are more computationally efficient than BDF2, and 
	ROW methods can potentially be more efficient than ESDIRK methods. However, the convergence tolerance of the GMRES solver for ROW methods needs to be sufficiently tight to preserve the nominal order of accuracy. In general, ESDIRK methods allow a larger physical time step size for unsteady flow simulation than ROW methods do. 
	
		\section*{Key Words}
		Linearly Implicit Rosenbrock-Wanner;  ESDIRK; Backward Differentiation Formula; Jacobian-Free; Flux Reconstruction/Correction Procedure via Reconstruction; High-Order Spatiotemporal Methods; Unsteady Flows
	
	\section{Introduction}
	High-order computational fluid dynamics (CFD) methods have been attracting much research attention in past decades due to their superior numerical properties that enable high-fidelity simulation of intricate flows. High-order spatial discretization methods, such as discontinuous Galerkin (DG) methods~\cite{CockburnShu89,bassi1997high,NDG08}, and flux reconstruction/correction procedure via reconstruction (FR/CPR) methods~\cite{Huynh2007, Huynh2009, ZJWang2009, Vincent2011}, have shown their capabilities of dealing with turbulent flows~\cite{Bassi2005,Liang2009,Persson11,bassi2016development,Fidkowski2016,Wang17,Vincent17,Marvriplis2018}. Usually, the high-order explicit strong stability preserving Runge-Kutta (SSPRK) methods~\cite{gottlieb2001strong} are used to integrate the semi-discretized governing equations for unsteady flow simulation. However, due to the Courant–Friedrichs–Lewy (CFL) number constraint, explicit time integration methods may not be the optimal choice for efficient numerical simulation of stiff flow problems, such as wall-bounded turbulent flows at high Reynolds numbers.  
	
	The implicit time integration methods have a better stability property than explicit ones. The backward differentiation formula (BDF) is very popular for its ease of implementation. However, BDF methods are not A-stable when the order of accuracy exceeds two. An alternative is the implicit Runge-Kutta (IRK) method~\cite{Butcher2002}. We note that fully coupled IRK methods are not widely used in the CFD community due to the complication of solving fully coupled nonlinear systems. Solution strategies based on the dual time stepping procedure~\cite{jameson1991time} have recently been reported by Jameson~\cite{jameson2017evaluation}. In that work, several fully coupled IRK methods have been evaluated for unsteady flow simulation. To decrease computational complexity, diagonally implicit Runge-Kutta (DIRK) and singly diagonally implicit Runge-Kutta (SDIRK) can be used.  A comprehensive review of DIRK methods by Kennedy and Carpenter can be found in Ref.~\cite{Kennedy2016}. A recent work by Vermeire and Vincent~\cite{vermeire2017behaviour} have analyzed dispersion and dissipation properties of fully-discrete high-order FR methods with two SDIRK schemes, and provided insights on their suitability for implicit large eddy simulation. As a special case of SDIRK, the explicit first stage singly diagonally implicit Runge-Kutta method (ESDIRK) reduces the degree of the nonlinear systems of SDIRK by one. Comparisons of BDF and ESDIRK methods have been performed in Refs.~\cite{bijl2002implicit, wang2007implicit}. ESDIRK methods are found to be more efficient than BDF methods. We note that for BDF and DIRK methods,  a nonlinear system needs to be solved at each step or each stage.
	
	Rosenbrock methods linearize the nonlinear equations in DIRK methods. Therefore, only one linear equation needs to be solved at each stage. Besides, the Jacobian matrix for this type of multistage methods only needs to be calculated once at the first stage. All these features can potentially make linearly implicit Rosenbrock methods more computationally efficient than DIRK methods. In traditional Rosenbrock methods, exact Jacobian matrices~\cite{Baker1988,lang2001ros3p} need to be calculated to ensure accuracy. In many stiff flow problems, analytical Jacobian is not easy to obtain; and matrix-based implementation can consume tremendous memory, thus impeding efficient simulation of large-scale flow problems. Rosenbrock-Wanner (ROW) methods ~\cite{rang2005new,rang2014analysis,steinebach1995order} can preserve the nominal order of accuracy with an approximate Jacobian matrix. 
	This flexibility makes it possible to implement a Jacobian-free Krylov subspace solver. The Rosenbrock-Krylov (ROK) methods~\cite{tranquilli2014rosenbrock} reformulate the Rosenbrock/ROW method such that stage vectors are obtained from the Krylov subspace using the modified Arnoldi iteration. The ROK methods naturally favor the matrix-free implementation. We notice that even stiffly accurate Rosenbrock methods can suffer from order reduction for moderately stiff problems and improvements have been developed in Ref.~\cite{rang2015improved}. 
	Recent research on the performance of Rosenbrock methods on solving the Navier-Stokes equations with high-order DG-type schemes have been reported in a series of works~\cite{Bassi2015,liu2016comparative,franciolini2017efficiency,wang2018parallel}.
	
	
	If Jacobian-free implementation is used, the advantage that the Jacobian matrix of Rosenbrock methods only needs to be evaluated once for each time step does not hold anymore. In fact, the performance of ROW methods and ESDIRK methods highly depends on how the linear and nonlinear systems are solved. Blom \textit{et al.}~\cite{blom2016comparison} have shown that ROW methods are not necessarily more efficient than ESDIRK methods when a second-order central finite volume method is used. Liu \textit{et al.}~\cite{liu2016comparative} has conducted a comparative study of some third-order ROW methods and a third-order ESDIRK (ESDIRK3)~\cite{bijl2002implicit} method with a third-order hierarchical WENO (weighted essentially non-oscillatory) reconstructed discontinuous
	Galerkin (rDG) method. They observed that the third-order ROW methods tested are more efficient than ESDIRK3. Sarshar \textit{et al.}~\cite{sarshar2017numerical} conducted a comparative study of various matrix-free implicit time-stepping methods on solving two-dimensional (2D) Navier-Stokes equations including SDIRK, Rosenbrock, ROW, and ROK using a second-order spatial discretization. It is found that ROK can be a competitive method compared to other implicit time integrations.
	
	\textit{Contributions}. In most numerical simulation of the unsteady Navier-Stokes equations with the method of lines, numerical dissipation and dispersion are primarily emphasized for spatial discretization methods, assuming that the influence of time integration methods on numerical simulation is negligible compared to that of spatial discretization. However, the assumption may not hold especially when large time steps, allowed by implicit time integrators, are employed. Several recent works~\cite{Yang2013,vermeire2017behaviour,ZJW2018} have confirmed that time integration methods can have significant impact on unsteady flow simulation with high-order spatial formulations. This motivates the present study to conduct a systematic comparison of accuracy and efficiency between several widely used implicit time integration methods, including Jacobian-free ROW (second-, third- and fourth-order schemes), ESDIRK (second-, third- and fourth-order schemes) and BDF2, for unsteady numerical simulation of stiff flow problems. In this study, the high-order FR methods are used to discretize the spatial domain. For ESDIRK and BDF2, the pseudo-transient continuation~\cite{jameson1991time} is used to solve the nonlinear systems. The restarted GMRES solver~\cite{Persson08} with the element-Jacobi preconditioner serves as the linear solver. The Jacobian-free implementation decreases the tremendous memory consumption of matrix-based implementation. The accuracy and efficiency of the implicit time integrators are compared using several laminar and turbulent flows, including 2D isentropic vortex propagation, 2D laminar flow over a cylinder, and three-dimensional (3D) Taylor-Green vortex evolution. 
	
	\textit{Article Organization}. The rest of the paper is organized as follows. Section~\ref{Section_GE} gives a brief introduction of the governing equations. The FR method is reviewed in Section~\ref{Section_FR}. Section~\ref{Section_TI} introduces all the time integration methods tested in this study. Numerical results and corresponding discussions are documented in Section~\ref{Section_NR}. We summarize this work in the last section. 
	
	\section{Governing Equations}\label{Section_GE}
	The 3D unsteady compressible Navier-Stokes equations can be written as 
	\begin{equation} \label{ns_equation}
	\frac{\partial \boldsymbol{q}}{\partial t} + \nabla \cdot \mathbb{F} = 0,
	\end{equation}
	where $ \boldsymbol{q} = (\rho, \rho u_j, E)^{T},\ j= 1,2,3,$ is the vector of conserved variables, and $\mathbb{F}$ is the corresponding flux tensor. Specifically, $\rho$ is the fluid density, $u_j, j=1,2,3,$ are the velocities in three orthogonal directions, and $ E=\frac{p}{\gamma-1}+\frac{1}{2} \rho \sum_{k=1}^{3} (u_k u_k) $ is the total energy per unit volume, where $ p=\rho RT $ is the pressure, $T$ is the temperature, $ R = C_{p} - C_{v} $ is the ideal gas constant, $\gamma$ is the specific heat ratio defined as $ \gamma = C_{p}/C_{v} $, and $C_{p}$ and $C_{v}$ are specific heat capacities at constant pressure and volume, respectively. In this study, $\gamma$ is set as 1.4. The flux tensor $ \mathbb{F} $  consists of the inviscid part and viscous part, which can be expressed as $ \mathbb{F} = \mathbb{F}_{inv}(\boldsymbol{q})-\mathbb{F}_{vis}(\boldsymbol{q},\nabla\boldsymbol{q}) $. Note that $\mathbb{F}_{inv}(\boldsymbol{q})$ and $\mathbb{F}_{vis}(\boldsymbol{q},\nabla\boldsymbol{q})$ can be rewritten in a vector format as: for any $j, \ j \in \{ 1,2,3 \},$ 
	\begin{equation}	
	\mathbb{F}_{inv,j}(\boldsymbol{q}) = 
	\begin{pmatrix}
	\rho u_j\\
	\rho u_1 u_j +\delta_{1j}p\\
	\rho u_2 u_j +\delta_{2j}p\\
	\rho u_3 u_j +\delta_{3j}p\\
	u_j(E+p)
	\end{pmatrix},\ \text{and} \
	\mathbb{F}_{vis,j}(\boldsymbol{q},\nabla \boldsymbol{q}) = 
	\begin{pmatrix}
	0\\
	\tau_{1j}\\
	\tau_{2j}\\
	\tau_{3j}\\
	\sum_{k=1}^{3} u_k\tau_{kj}-K_j
	\end{pmatrix},
	\end{equation}
	where $\delta_{ij}$ is the Kronecker delta, $\tau_{ij}$ is the viscous stress defined as 
	\begin{equation*}
	\tau_{ij} = \mu \left(\frac{\partial u_i}{\partial x_j}+\frac{\partial u_j}{\partial x_i}\right)-\frac{2}{3} \mu \sum_{k=1}^{3}\frac{\partial u_k}{\partial x_k}\delta_{ij} \ \text{with} \ i = 1, 2, 3,
	\end{equation*}
	and following Fourier's law, the heat flux $K_j$ is defined as
	$K_j = -\kappa \partial T / \partial x_j$.
	In this study, the thermal conductivity $\kappa$ is calculated from the Prandtl number $Pr$ as $ \kappa=\mu C_p/Pr $. The fluid viscosity $ \mu  $ and the Prandtl number $Pr$ are treated as constants.	
	\section{Flux Reconstruction Methods}\label{Section_FR}
	To achieve an efficient implementation of Eq.~\eqref{ns_equation}, we transfer the Navier-Stokes equations from the physical domain $ (x,y,z) $ to the computational domain $ (\xi,\eta, \zeta) $. Thus Eq.~\eqref{ns_equation} can be expressed as 
	
	\begin{equation}\label{transfered_ns_equation}
	\frac{\partial \boldsymbol{Q}}{\partial t} + \frac{\partial \boldsymbol{F}}{\partial \xi} + \frac{\partial \boldsymbol{G}}{\partial \eta} + \frac{\partial \boldsymbol{H}}{\partial \zeta} = 0,
	\end{equation}
	where 
	\begin{equation}
	\begin{cases}\label{trans_detail}
	\boldsymbol{Q} = |\boldsymbol{J}|\boldsymbol{q}, \\
	\boldsymbol{F} = |\boldsymbol{J}|(\boldsymbol{f}\xi_{x}+\boldsymbol{g}\xi_{y}+\boldsymbol{h}\xi_{z}),\\
	\boldsymbol{G} = |\boldsymbol{J}|(\boldsymbol{f}\eta_{x}+\boldsymbol{g}\eta_{y}+\boldsymbol{h}\eta_{z}),\\
	\boldsymbol{H} = |\boldsymbol{J}|(\boldsymbol{f}\zeta_{x}+\boldsymbol{g}\zeta_{y}+\boldsymbol{h}\zeta_{z}),
	\end{cases}
	\end{equation}
	and 
	\begin{equation}\label{jacobi_info}
	\boldsymbol{J} = \frac{\partial(x,y,z)}{\partial(\xi,\eta,\zeta)},\ \text{and}\ |\boldsymbol{J}| = \det(\boldsymbol{J}).
	\end{equation}
	Herein, $\boldsymbol{f} = \mathbb{F}_{1}$, $\boldsymbol{g} = \mathbb{F}_{2}$, and $\boldsymbol{h} = \mathbb{F}_{3}$ are flux vectors in the $x$, $y$, and $z$ directions of a Cartesian coordinate system. ${\partial \xi_{i}}/{\partial x_{j}}, \ i,j=1,2,3,$ are metrics in the coordinate transformation.
	
	The flux polynomial reconstructed from the FR method consists of two parts, one of which is the local flux polynomial and the other is the correction polynomial. On solving Eq.~\eqref{transfered_ns_equation}, the reconstructed polynomials $ \widetilde{\boldsymbol{F}} $, $ \widetilde{\boldsymbol{G}} $ and $ \widetilde{\boldsymbol{H}}$ of $ \boldsymbol{F} $, $\boldsymbol{G} $ and $ \boldsymbol{H} $ can be expressed as
	\begin{equation}
	\begin{cases}
	\widetilde{\boldsymbol{F}} = \boldsymbol{F}^l+\boldsymbol{F}^c,\\
	\widetilde{\boldsymbol{G}} = \boldsymbol{G}^l+\boldsymbol{G}^c,\\
	\widetilde{\boldsymbol{H}} = \boldsymbol{H}^l+\boldsymbol{H}^c,
	\end{cases}
	\end{equation}
	where the superscript `$l$' stands for the local flux and `$c$' stands for the correction flux. Consequently, Eq.~\eqref{ns_equation} can be rewritten as
	\begin{equation}\label{discrete_ns}
	\frac{\partial \boldsymbol{q}}{\partial t} + \frac{\partial \boldsymbol{f}^l}{\partial x} +   \frac{\partial \boldsymbol{g}^l}{\partial y} + \frac{\partial \boldsymbol{h}^l}{\partial z} + \frac{1}{|\boldsymbol{J}|}\left(\frac{\partial {\boldsymbol{F}}^c}{\partial \xi}+\frac{\partial {\boldsymbol{G}}^c}{\partial \eta}+\frac{\partial {\boldsymbol{H}}^c}{\partial \zeta}\right)= 0
	\end{equation}
	
	For hexahedral elements, $ \boldsymbol{F}^c $ , $ \boldsymbol{G}^c $ and $ \boldsymbol{H}^c $ can be explicitly expressed as 
	\begin{equation}
	\begin{cases}
	\boldsymbol{F}^c(\xi,\eta,\zeta) = (\widetilde{\boldsymbol{F}}(-1,\eta,\zeta)-\boldsymbol{F}^{l}(-1,\eta,\zeta))g_L(\xi)
	+(\widetilde{\boldsymbol{F}}(1,\eta,\zeta)-\boldsymbol{F}^{l}(1,\eta,\zeta))g_R(\xi),\\
	\boldsymbol{G}^c(\xi,\eta,\zeta) = (\widetilde{\boldsymbol{G}}(\xi,-1,\zeta)-\boldsymbol{G}^{l}(\xi,-1,\zeta))g_L(\eta)+(\widetilde{\boldsymbol{G}}(\xi,1,\zeta)-\boldsymbol{G}^{l}(\xi,1,\zeta))g_R(\eta),\\
	\boldsymbol{H}^c(\xi,\eta,\zeta) = (\widetilde{\boldsymbol{H}}(\xi,\eta,-1)-\boldsymbol{H}^{l}(\xi,\eta,-1))g_L(\zeta)+(\widetilde{\boldsymbol{H}}(\xi,\eta,1)-\boldsymbol{H}^{l}(\xi,\eta,1))g_R(\zeta),
	\end{cases}
	\end{equation}
	where $ g_{L/R} $ are the correction polynomials. In this study, we employ the Radau polynomials to recover the nodal FR-DG method. $ \widetilde{\boldsymbol{F}} $, $ \widetilde{\boldsymbol{G}} $  and $ \widetilde{\boldsymbol{H}} $ at element interfaces are referred as numerical fluxes $ \boldsymbol{F}^{num} $, $ \boldsymbol{G}^{num} $ and $ \boldsymbol{H}^{num} $.  The inviscid common fluxes can be obtained from approximate Riemann solvers. In this study, the Roe approximate Riemann solver~\cite{Roe81} is used to calculate the common fluxes at the cell interfaces in their normal directions as
	\begin{equation}\label{invis_com}
	\boldsymbol{f}_{\boldsymbol{n},inv}^{com} = \frac{\boldsymbol{f}_{\boldsymbol{n},inv}^{+}+\boldsymbol{f}_{\boldsymbol{n},inv}^{-}}{2}-\boldsymbol{R}|\Lambda|\boldsymbol{R}^{-1}\frac{\boldsymbol{q}^{+}-\boldsymbol{q}^{-}}{2},
	\end{equation}
	where superscripts `$-$' and `$+$' denote the left of right side of the current interface, the subscript $ \boldsymbol{n} $ is the unit normal direction from left to right, $\boldsymbol{f}_{\boldsymbol{n}} = f n_x + g n_y + h n_z$ is the flux projection in the $ \boldsymbol{n} $ direction, $\Lambda$ is a diagonal matrix consisting of the eigenvalues of the Jacobian $\partial \boldsymbol{f}_{\boldsymbol{n}} /  \partial \boldsymbol{q}$, and $\boldsymbol{R}$ consists of the corresponding right eigenvectors evaluated with the Roe-averaged variables. Numerical common fluxes can be obtained as 
	\begin{equation}\label{norm_num_f}
	\begin{cases}
	\boldsymbol{F}^{num}={|\boldsymbol{J}|}|\nabla\xi|\boldsymbol{f}^{com}_{\boldsymbol{n}} \sign(\boldsymbol{n}\cdot\nabla\xi),\\
	\boldsymbol{G}^{num}={|\boldsymbol{J}|}|\nabla\eta|\boldsymbol{f}^{com}_{\boldsymbol{n}} \sign(\boldsymbol{n}\cdot\nabla\eta),\\
	\boldsymbol{H}^{num}={|\boldsymbol{J}|}|\nabla\zeta|\boldsymbol{f}^{com}_{\boldsymbol{n}} \sign(\boldsymbol{n}\cdot\nabla\zeta).
	\end{cases}
	\end{equation}
	The common viscous fluxes at the cell interfaces are $ \boldsymbol{f}_{n,vis}^{com} = \boldsymbol{f}_{vis}(\boldsymbol{q}^{+},\nabla \boldsymbol{q}^{+},\boldsymbol{q}^{-},\nabla \boldsymbol{q}^{-}) $. Here we need to define the common solution $ \boldsymbol{q}^{com} $ and common gradient $ \nabla \boldsymbol{q}^{com} $ at the cell interface. 
	By simply taking average of the primitive variables, we get 
	\begin{equation}\label{com_q}
	\boldsymbol{q}^{com} = \frac{\boldsymbol{q}^{+}+\boldsymbol{q}^{-}}{2}.
	\end{equation}
	The common gradient is computed as 
	\begin{equation}\label{com_grad_q}
	\nabla \boldsymbol{q}^{com} = \frac{\nabla\boldsymbol{q}^{+}+\boldsymbol{r}^{+}+\nabla\boldsymbol{q}^{-}+\boldsymbol{r}^{-}}{2},
	\end{equation}
	where $ \boldsymbol{r}^{+} $ and $ \boldsymbol{r}^{-} $ are the corrections to the gradients on the interface. The second
	approach of Bassi and Rebay (BR2)~\cite{Bassi2005} is used to calculate the corrections. For the hexahedral element, the correction terms are defined as
	\begin{equation}
	\boldsymbol{r} = \gamma (\boldsymbol{q}^{com}-\boldsymbol{q}_{L/R})\boldsymbol{n},\ \text{and} \ \gamma = |\nabla \varpi|g'_{L/R}(\varpi)\sign(\boldsymbol{n}\cdot\nabla\varpi),
	\end{equation}
	where $ \varpi \in \{\xi, \eta, \zeta \} $, and $ \boldsymbol{q}_{L/R} $ is the local solution on the interface. If the interface is the left boundary of the element, then the local solution $ \boldsymbol{q}_L $ is used, and $ \varpi $ is $ -1 $; if the interface is the right boundary of the element, then the local solution $ \boldsymbol{q}_R $ is used, and $ \varpi $ is $ 1 $. In this study,
	$ g(\pm1)=\pm(P+1)(P+2)/2 $ is used to stabilize the FR method ~\cite{gao2013differential}, where $P$ is the polynomial degree of the solution.

	\section{Time Integration Methods}\label{Section_TI}
	To solve Eq.~\eqref{discrete_ns}, we rewrite this equation in a semi-discretized form as 
	\begin{equation}\label{simplified_ode}
	\frac{\partial \boldsymbol{q}}{\partial t} = \boldsymbol{R}(\boldsymbol{q}),
	\end{equation}
	for the convenience of notation. We use the method of lines approach to solve this equation: the partial differential equation is first discretized in space with the FR formulation in Section~\ref{Section_FR}, and then is marched in time with the time integrators described below.
	\subsection{The backward differentiation formula}
	The BDF method for solving Eq.~\eqref{simplified_ode} can be expressed as 
	\begin{equation}\label{BDF}
	\boldsymbol{q}^{n+1} = \Delta t\omega \boldsymbol{R}(\boldsymbol{q}^{n+1})+\sum_{j=1}^{s}a_j\boldsymbol{q}^{n+1-j},\\
	\end{equation}
	BDF  methods are not A-stable when the order of accuracy exceeds two. In this study, we only consider the BDF2 method for comparison. For BDF2, $ s=2 $, $ \omega = 2/3 $, $ a_1=4/3 $ and $ a_2=-1/3 $.  
	\subsection{The explicit first stage, single diagonally implicit Runge-Kutta method }
	The ESDIRK methods can be written as 
	\begin{equation}\label{ESDIRK}
	\begin{cases}
	\boldsymbol{q}^{n+1} = \boldsymbol{q}^{n}+\Delta t \sum_{i=1}^{s}b_i\boldsymbol{R}(\boldsymbol{q}_i),\\
	\boldsymbol{q}_i = \boldsymbol{q}^n+\Delta t \sum_{j=1}^{i}a_{ij}\boldsymbol{R}(\boldsymbol{q}_j),i=1,\dots,s,	
	\end{cases}	
	\end{equation}
	where $ s $ is the number of stages. The second-order, three-stage ESDIRK2~\cite{Kennedy2016}, third-order, four-stage ESDIRK3~\cite{bijl2002implicit} and fourth-order, six-stage ESDIRK4~\cite{bijl2002implicit} methods are studied in this paper. 
	All the ESDIRK methods investigated in this study have the feature that 
	\begin{equation}
	a_{ii} = \begin{cases}
	0,&i=1,\\
	\omega, & i \neq 1.
	\end{cases}
	\end{equation}
	Therefore, we can rewrite Eq.~\eqref{ESDIRK} as 
	\begin{equation}\label{ESDIRK2}
	\begin{cases}
	\boldsymbol{q}^{n+1} = \boldsymbol{q}^{n}+\Delta t
	\sum_{i=1}^{s}b_i\boldsymbol{R}(\boldsymbol{q}_i),\\
	\boldsymbol{q}_1 = \boldsymbol{q}^n,\\
	\boldsymbol{q}_i = \Delta t \omega \boldsymbol{R}(\boldsymbol{q}_i)+\boldsymbol{q}^n+\Delta t \sum_{j=1}^{i-1}a_{ij}\boldsymbol{R}(\boldsymbol{q}_j),i=2,\dots,s.	
	\end{cases}	
	\end{equation}

	\subsection{Linearly implicit Rosenbrock methods}
		
	The general form of Rosenbrock methods for solving Eq.~\eqref{simplified_ode} from time step $ n $ to $ n+1 $ can be written as
	\begin{equation}\label{Rosenbrock_General}
	\begin{cases}
	\boldsymbol{q}^{n+1} = \boldsymbol{q}^n+\sum_{j=1}^{s}m_j\boldsymbol{Y}_j,\\	
	\left(\frac{\boldsymbol{I}}{\omega \Delta t}-\frac{\partial \boldsymbol{R}}{\partial\boldsymbol{q}}\right)^n\boldsymbol{Y}_i = \boldsymbol{R}\left(\boldsymbol{q}^n+\sum_{j=1}^{i-1}a_{ij}\boldsymbol{Y}_j\right)+\frac{1}{\Delta t}\sum_{j=1}^{i-1}c_{ij}\boldsymbol{Y}_j,\ i=1,2,\ldots,s,
	\end{cases}
	\end{equation}
	where $ s $ is the number of stages. 
	Rosenbrock-Wanner methods have the flexibility of evaluating the Jacobian matrix approximately while preserving the accuracy of the schemes. In this study, we restrict our attention to three popular stiffly accurate Rosenbrock-Wanner methods, i.e., the second-order, three-stage ROS2PR~\cite{rang2014analysis}, the third-order, four-stage ROS34PW2~\cite{rang2005new} and the fourth-order, six-stage RODASP~\cite{steinebach1995order}. We note that it has been shown in Ref.~\cite{rang2015improved} that ROS34PW2 and RODASP can suffer from order reduction.  To simplify the notation, we use ROW2, ROW3 and ROW4 to denote them, respectively.
	
	\subsection{Iterative methods}
	The nonlinear equations in Eq.~\eqref{BDF} and Eq.~\eqref{ESDIRK2} can be written as 
	\begin{equation} \label{eq:nonlinear}
	 \boldsymbol{F}(\boldsymbol{q}^*) = 0, 
	\end{equation}
	where $ \boldsymbol{q}^{*} $ is $ \boldsymbol{q}^{n+1} $ for BDF and $ \boldsymbol{q}^{i} $ for ESDIRK, respectively. For BDF methods,
	\begin{equation}\label{BDF_newton}
	\boldsymbol{F}(\boldsymbol{q}^{n+1}) = \left(\frac{1}{\omega \Delta t}\boldsymbol{q}^{n+1} - \boldsymbol{R}(\boldsymbol{q}^{n+1})\right)-\frac{1}{\omega\Delta t}\sum_{j=1}^{s}a_j\boldsymbol{q}^{n+1-j}.\\
	\end{equation}
	For ESDIRK methods, 
	\begin{equation}\label{SDIRK_newton}
	\boldsymbol{F}(\boldsymbol{q}^{i}) = \left(\frac{1}{\omega\Delta t}\boldsymbol{q}_i -\boldsymbol{R}(\boldsymbol{q}_i)\right)-\frac{1}{\omega\Delta t}\left(\boldsymbol{q}^n+\Delta t\sum_{j=1}^{i-1}a_{ij}\boldsymbol{R}(\boldsymbol{q}_j)\right),i=2,\dots,s.
	\end{equation}

	In this work, the pseudo-transient continuation is employed to solve the nonlinear equations~\cite{jameson1991time,bijl2002implicit,jameson2017evaluation}.
	The pseudo-transient continuation is an alternative of inexact Newton's method to solve the steady state equation $ \boldsymbol{F}(\boldsymbol{q}^*) = 0 $ iteratively as 
	\begin{equation}\label{pseudo_transient}
	\frac{\boldsymbol{q}^{k+1,*}-\boldsymbol{q}^{k,*}}{\Delta \tau}=-\boldsymbol{F}(\boldsymbol{q}^{k+1,*}),
	\end{equation}
	Herein, $ k $ is the iteration step for the pseudo-transient continuation. Eq.~\eqref{pseudo_transient} can be linearized as 
	\begin{equation}\label{linearized_pseudo_transient}
	\left(\frac{\boldsymbol{I}}{\Delta \tau} + \frac{\partial \boldsymbol{F}}{\partial \boldsymbol{q}}\right)^k\Delta \boldsymbol{q}^{k,* }= -\boldsymbol{F}(\boldsymbol{q}^{k,*}).
	\end{equation}
	For a steady problem, as $ k \to \infty $, $ \tau \to \infty $ and $ \Delta \boldsymbol{q}^k \to 0 $. Therefore, $ \boldsymbol{F}(\boldsymbol{q}^{k,*})\to \boldsymbol{F}(\boldsymbol{q}^{*}) $. The Jacobian matrix for fully implicit methods can be expressed as 
	\begin{equation}
	\left(\frac{\partial \boldsymbol{F}}{\partial \boldsymbol{q}}\right)^k=\left(\frac{\boldsymbol{I}}{\omega \Delta t}-\frac{\partial \boldsymbol{R}}{\partial\boldsymbol{q}}\right)^k.
	\end{equation}
	Substitute the Jacobian matrix into Eq.~\eqref{linearized_pseudo_transient} to obtain the final form of the linear system in the pseudo-transient continuation procedure as follows
	\begin{equation}\label{linearized_pseudo_transient_final}
	\left(\frac{\boldsymbol{I}}{\Delta \tau} +\frac{\boldsymbol{I}}{\omega \Delta t}-\frac{\partial \boldsymbol{R}}{\partial\boldsymbol{q}}\right)^k\Delta \boldsymbol{q}^{k,* }= -\boldsymbol{F}(\boldsymbol{q}^{k,*}).
	\end{equation}
	The pseudo-transient continuation procedure requires an adaptation algorithm of the pseudo time step size to complete the method. In this study, we employ successive evolution relaxation (SER) algorithm~\cite{mulder1985experiments} as  
	\begin{equation}
	\tau^{0} = \tau_{init},
	\Delta \tau^{k+1} = \min\left(\Delta \tau^{k}\frac{||\boldsymbol{F}||_{L_2}^{k-1}}{||\boldsymbol{F}||_{L_2}^{k}}, \Delta \tau_{max} \right).
	\end{equation}
	As the pseudo time marches forward, a series of linear equations~\eqref{linearized_pseudo_transient_final} are successively solved until convergence. 
	Ideally, we would expect that when $ \Delta \tau $ approaches $ \Delta \tau_{max}=\infty $, the residual of the pseudo-transient procedure will gradually converge to machine zero. However, in our practice of simulating wall-bounded flows with grids  clustered in near wall regions, we have to choose a moderately large $ \Delta \tau_{max} $ to ensure that the residual of the linear solver (e.g., the restarted GMRES as will be explained in the next subsection) can at least drop by one magnitude when $ \Delta \tau $ equals to $ \Delta \tau_{max} $. Once the linear solver fails, we would reject the current pseudo-transient continuation iteration and decrease $ \Delta \tau_{max} $ by half to continue. In this study, if not specifically mentioned, $ \Delta \tau_{init} = \Delta t $ and $ \Delta \tau_{max} = 10^{20}  $.
	
	\subsection{Jacobian-free implementation of GMRES}
	The restarted GMRES is employed to solve Eq.~\eqref{linearized_pseudo_transient_final} as well as the linear system in Eq.\eqref{Rosenbrock_General}. All linear systems can be expressed as 
	\begin{equation} \label{eq:linear}
	AX=b,
	\end{equation}
	where 
	\begin{equation}
	A=D(\Delta t, \Delta \tau)-\frac{\partial R}{\partial q}.
	\end{equation}
	The first term $ D(\Delta t, \Delta \tau) $ on the right-hand side of the above equation is a diagonal matrix related to $ \Delta t $ and $ \Delta \tau $.
	The GMRES method approximates the exact solution by a vector $ x_n\in K_n $ that minimizes the Euclidean norm $ ||Ax_n-b|| $ where $ K_n $ is the $ n $-th Krylov subspace 
	\begin{equation}
	K_n = \text{span}\{b,Ab,A^2b,\cdots,A^{n-1}b\}.
	\end{equation}
	In the GMRES solver, the sparse matrix $ \frac{\partial \boldsymbol{R}}{\partial\boldsymbol{q}} $ in $ A $ only appears in the matrix-vector product. For an unknown vector $ \boldsymbol{X} $, the matrix-vector product
	can be approximated as 
	\begin{equation}
	\left(\frac{\partial \boldsymbol{R}}{\partial\boldsymbol{q}}\right)\boldsymbol{X} = \frac{\boldsymbol{R}(\boldsymbol{q}+\epsilon \boldsymbol{X})-\boldsymbol{R}(\boldsymbol{q})}{\epsilon}+O(\epsilon).
	\end{equation}
	$ \epsilon = 10^{-8} $ in this study.
	Interested readers are referred to Ref.~\cite{knoll2004jacobian} for more discussions on this approximation.
	
	The performance of the Newton-Krylov methods substantially depends on the preconditioner. In the context of Jacobian-free implementation, the diagonal blocks of the Jacobian matrix, i.e., the element-Jacobi preconditioner, can be an effective choice among low-storage preconditioners, such as the matrix-free LU-SGS (Lower-Upper Symmetric-Gauss-Seidel) preconditioner~\cite{sharov2000implementation}, and $ p $-multigrid preconditioner~\cite{franciolini2018p}. In this study, only the element-Jacobi preconditioner is considered. The restart number of the GMRES solver is set as 60.

	\section{Numerical Results}\label{Section_NR}
	All the simulations presented in this section are performed on the High Performance Computing Facility (HPCF) of the University of Maryland, Baltimore county (UMBC). All nodes used in simulations have two Intel E5-2650v2 Ivy Bridge (2.6 GHz, 20 MB cache) processors with eight cores apiece, for a total of 16 cores per node. A quad data rate (QDR) Infiniband switch connects all the nodes. Ideally the system can achieve a latency of 1.2 $\mu$sec to transfer a message between two nodes, and can support a bandwidth of up to 40 gigabits per second (40Gbps). Every node possesses 64 GB RAM. All nodes are running Red Hat Enterprise Linux 6.4. We employ g++ (GCC) 4.8.4 with mpich-3.1.4 to compile the code for parallel simulation.
	\subsection{Vortex propagation}
	The 2D isentropic vortex propagation problem is employed as a benchmark test test case to investigate the accuracy and efficiency of different time integrators in this subsection. The free stream flow conditions are set as $ (\rho,u,v,Ma) = (1,1,1,0.5) $ and the ideal gas constant $ R $ is set as $1.0$ for this case. The perturbation is defined as~\cite{Bassi2015}
	\begin{equation}\label{VP_Solution}
	\begin{cases}
	\delta u=-\frac{\alpha}{2\pi}(y-y_0)e^{\phi(1-r^2)},\\
	\delta v=\frac{\alpha}{2\pi}(x-x_0)e^{\phi(1-r^2)},\\
	\delta T=-\frac{\alpha^2(\gamma-1)}{16\phi\gamma\pi^2} e^{2\phi(1-r^2)},\\
	dS=0,\\
	\end{cases}	
	\end{equation}
	where $ \phi = \frac{1}{2}$ and $ \alpha = 5 $ are parameters that define the vortex strength, and $ r=(x-x_0)^{2}+(y-y_0)^{2} $ is the distance to the center of the vortex $ (x_0,y_0) = (0,0) $ at $ t=0 $. The periodic domain is defined as $ \Omega = [-10,10]^2 $. We simulate this problem on a uniform mesh of $ 50\times50 $ elements with the $ P^6 $ (i.e. $7th$ order) FR method  for one period. The time step size $ \Delta t $ is refined from $ T/100 $ to $ T/800 $. The error of any variable $ s $ is defined as 
	\begin{equation}
	Error(s) = \sqrt{\frac{\int_\Omega(s_{exact}-s_{num})^2dV}{V}},
	\end{equation}
	where $s_{exact}$ is the exact value, $s_{num}$ is the numerical value, and $V$ is the volume of the domain $ \Omega $.
	
	To monitor the convergence of the nonlinear system~\eqref{eq:nonlinear} and linear system~\eqref{eq:linear}, two convergence criteria are set up: one is for pseudo time iterations, denoted as $tol_{rel,nonlinear}$; the other is for the restarted GMRES linear solver, denoted as $tol_{rel,linear}$. Herein, the subscript `$rel$' stands for `relative'. This is due to that the relative residual with respect to that at the first step (i.e., the residual at the first step in pseudo time iterations for a nonlinear system, and the residual at the first step in the restarted GMRES for a linear system) is used in this study.  
	
	\subsubsection{Effect of pseudo-transient continuation convergence criterion and GMRES convergence criterion on ESDIRK and BDF2}	
	The pseudo-transient continuation does not require exact solution of the linear system at each iteration. We first employ a relative big tolerance $ tol_{rel,linear} =10^{-1}$ for GMRES to investigate the effect of $ tol_{rel,nonlinear} $ on accuracy and efficiency of fully implicit methods. The convergence study of ESDIRK and BDF2 of different $ tol_{rel,nonlinear} $, i.e., $ 10^{-2} $, $ 10^{-4} $, $ 10^{-6} $ and $ 10^{-8} $, is presented in Table~\ref{fullyimplicit_nonlinear}. We observe that when $ tol_{rel,nonlinear} $ is refined from $ 10^{-2} $ to $ 10^{-4} $, all ESDIRK methods studied here will preserve the nominal order of accuracy except that ESDIRK4 shows slightly order reduction  when $ \Delta t $ is refined from $ T/400 $ to $ T/800 $. ESDIRK methods have shown better accuracy and efficiency than BDF2. Generally, the higher the order of accuracy of ESDIRK is, the more efficient ESDIRK is (see Figure~\ref{vortex_convergence_all}). 
	
	We have also conducted a study on the effect of $ tol_{rel,linear} $ on efficiency when a small enough $ tol_{rel,nonlinear} = 10^{-6}$ is employed. $ tol_{rel,linear} $ spans in $ \{10^{-1}, 10^{-2}, 10^{-4}, 10^{-6}\} $. The timing results are presented in Table~\ref{fullyimplicit_linear}. For this specific problem, all values of $ tol_{rel,linear} $ lead to the same numerical errors as expected since the accuracy of ESDIRK and BDF2 methods solely depends on the convergence of pseudo-transient  continuation. As documented in Table~\ref{fullyimplicit_linear}, the computational cost will keep on increasing when we refine the tolerance $ tol_{rel,linear} $. This indicates that in order to save computational cost, $tol_{rel,linear}$ for ESDIRK and BDF2 needs not to be tight.

	\clearpage
   	\begin{center}
   		\begin{longtable}{|r|r|r|r|r|r|r|}
   			\caption{The convergence study for ESDIRK and BDF2 with different $ tol_{rel,nonlinear} $ when $ tol_{rel,linear} $ is set to $10^{-1}$.}	\label{fullyimplicit_nonlinear}	\\	
   			
   			\endfirsthead
   			
   			\hline
   			\multicolumn{7}{|l|}{Continuation of Table~\ref{fullyimplicit_nonlinear},}\\
   			\hline
   			\endhead
   			
   			\multicolumn{7}{|r|}{To be continued on next page.}\\
   			\hline
   			\endfoot
   			
   			\multicolumn{7}{| r |}{End of Table~\ref{fullyimplicit_nonlinear}.}\\
   			\hline
   			\endlastfoot

   			\hline
   			\multicolumn{7}{|c|}{$ tol_{rel,nonlinear}=10^{-2} $, $ tol_{rel,linear}=10^{-1} $}\\
   			\hline	
   			&$ \Delta t $ &  $ E_{L_2}( \rho )  $  & order  & $ E_{L_2}( u )  $  & order & CPU time(s)\\
   			\hline
   			\multirow{4}{2cm}{ESDIRK2}
   			&T/100 &$ 4.4097\times 10^{-4} $&     & $ 5.7334\times 10^{-3} $&     & 317\\
   			&T/200 &$ 1.1313\times 10^{-4} $& 1.97& $ 1.4622\times 10^{-3} $&1.97 & 603\\
   			&T/400 &$ 2.8315\times 10^{-5} $& 2.00& $ 3.6523\times 10^{-4} $&2.00 & 1159\\
   			&T/800 &$ 7.0714\times 10^{-6} $& 2.00& $ 9.1104\times 10^{-5} $&2.00 & 2247\\
   			\hline
   			\multirow{4}{2cm}{ESDIRK3}
   			&T/100 &$ 1.5341\times 10^{-4} $&     & $ 1.7108\times 10^{-3} $&    & 398\\
   			&T/200 &$ 2.3518\times 10^{-5} $& 2.71& $ 2.6017\times 10^{-4} $&2.90& 734\\
   			&T/400 &$ 3.1802\times 10^{-6} $& 2.89& $ 3.4016\times 10^{-5} $&2.94& 1392\\
   			&T/800 &$ 1.0661\times 10^{-6} $& 1.58& $ 9.7414\times 10^{-6} $&1.80& 2478\\
   			\hline
   			\multirow{4}{2cm}{ESDIRK4}
   			&T/100 &$ 4.5157\times 10^{-6} $&     & $ 3.2613\times 10^{-5} $&    & 425\\
   			&T/200 &$ 7.7894\times 10^{-7} $& 2.54& $ 4.7845\times 10^{-5} $&2.77& 773\\
   			&T/400 &$ 6.0568\times 10^{-8} $& 3.68& $ 3.2420\times 10^{-7} $&3.88& 1511\\
   			&T/800 &$ 2.2122\times 10^{-8} $& 1.45& $ 1.5485\times 10^{-7} $&1.07& 2802\\
   			\hline
   			\multirow{4}{2cm}{BDF2}
   			&T/100 &$ 1.9170\times 10^{-3} $&     & $ 3.0021\times 10^{-2} $&    & 292\\
   			&T/200 &$ 8.1561\times 10^{-3} $& 1.23& $ 1.1195\times 10^{-2} $&1.42& 548\\
   			&T/400 &$ 2.1372\times 10^{-4} $& 1.93& $ 2.8983\times 10^{-3} $&1.95& 1074\\
   			&T/800 &$ 5.6824\times 10^{-5} $& 1.91& $ 7.3097\times 10^{-4} $&1.99& 2133\\
   			\hline
   			\multicolumn{7}{|c|}{$ tol_{rel,nonlinear}=10^{-4} $, $ tol_{rel,linear}=10^{-1} $}\\
   			\hline	
   			&$ \Delta t $ &  $ E_{L_2}( \rho )  $  & order  & $ E_{L_2}( u )  $  & order & CPU time(s)\\
   			\hline
   			\multirow{4}{2cm}{ESDIRK2}
   			&T/100 &$ 4.4088\times 10^{-4} $&     & $ 5.7335\times 10^{-3} $&     & 375\\
   			&T/200 &$ 1.1278\times 10^{-4} $& 1.97& $ 1.4549\times 10^{-3} $&1.98 & 723\\
   			&T/400 &$ 2.8267\times 10^{-5} $& 2.00& $ 3.6420\times 10^{-4} $&2.00 & 1323\\
   			&T/800 &$ 7.0691\times 10^{-6} $& 2.00& $ 9.1040\times 10^{-5} $&2.00 & 2504\\
   			\hline
   			\multirow{4}{2cm}{ESDIRK3}
   			&T/100 &$ 1.5167\times 10^{-4} $&     & $ 1.6911\times 10^{-3} $&    & 537\\
   			&T/200 &$ 2.3024\times 10^{-5} $& 2.72& $ 2.5401\times 10^{-4} $&2.74& 956\\
   			&T/400 &$ 3.0921\times 10^{-6} $& 2.90& $ 3.3171\times 10^{-5} $&2.94& 1685\\
   			&T/800 &$ 3.9376\times 10^{-7} $& 2.97& $ 4.1908\times 10^{-6} $&2.98& 2890\\
   			\hline
   			\multirow{4}{2cm}{ESDIRK4}
   			&T/100 &$ 3.4484\times 10^{-6} $&     & $ 3.1216\times 10^{-5} $&    & 597\\
   			&T/200 &$ 2.1713\times 10^{-7} $& 3.99& $ 1.9593\times 10^{-6} $&4.00& 1046\\
   			&T/400 &$ 1.3573\times 10^{-8} $& 4.00& $ 1.2257\times 10^{-7} $&4.00& 1920\\
   			&T/800 &$ 1.1691\times 10^{-9} $& 3.54& $ 7.6729\times 10^{-9} $&3.81& 3193\\
   			\hline
   			\multirow{4}{2cm}{BDF2}
   			&T/100 &$ 1.9126\times 10^{-3} $&     & $ 3.0095\times 10^{-2} $&    & 329\\
   			&T/200 &$ 8.4137\times 10^{-4} $& 1.18& $ 1.1454\times 10^{-2} $&1.39& 627\\
   			&T/400 &$ 2.3001\times 10^{-4} $& 1.87& $ 2.9892\times 10^{-3} $&1.94& 1179\\
   			&T/800 &$ 5.8417\times 10^{-5} $& 1.98& $ 7.5360\times 10^{-4} $&1.99& 2294\\
   			\hline

   			\pagebreak
   			\multicolumn{7}{|c|}{$ tol_{rel,nonlinear}=10^{-6} $, $ tol_{rel,linear}=10^{-1} $}\\
   			\hline	
   			&$ \Delta t $ &  $ E_{L_2}( \rho )  $  & order  & $ E_{L_2}( u )  $  & order & CPU time(s)\\
   			\hline
   			\multirow{4}{2cm}{ESDIRK2}
   			&T/100 &$ 4.4088\times 10^{-4} $&     & $ 5.7335\times 10^{-3} $&     & 464\\
   			&T/200 &$ 1.1278\times 10^{-4} $& 1.97& $ 1.4549\times 10^{-3} $&1.98 & 814\\
   			&T/400 &$ 2.8267\times 10^{-5} $& 2.00& $ 3.6420\times 10^{-4} $&2.00 & 1503\\
   			&T/800 &$ 7.0691\times 10^{-6} $& 2.00& $ 9.1041\times 10^{-5} $&2.00 & 2672\\
   			\hline
   			\multirow{4}{2cm}{ESDIRK3}
   			&T/100 &$ 1.5165\times 10^{-4} $&     & $ 1.6911\times 10^{-3} $&    & 692\\
   			&T/200 &$ 2.3021\times 10^{-5} $& 2.72& $ 2.5399\times 10^{-4} $&2.74& 1138\\
   			&T/400 &$ 3.0914\times 10^{-6} $& 2.90& $ 3.3165\times 10^{-5} $&2.94& 1877\\
   			&T/800 &$ 3.9326\times 10^{-7} $& 2.97& $ 4.1867\times 10^{-5} $&2.99& 3369\\
   			\hline
   			\multirow{4}{2cm}{ESDIRK4}
   			&T/100 &$ 3.4502\times 10^{-6} $&     & $ 3.1249\times 10^{-5} $&    & 780\\
   			&T/200 &$ 2.1706\times 10^{-7} $& 3.99& $ 1.9593\times 10^{-6} $&4.00& 1310\\
   			&T/400 &$ 1.3584\times 10^{-8} $& 4.00& $ 1.2270\times 10^{-7} $&4.00& 2156\\
   			&T/800 &$ 8.4961\times 10^{-10} $& 4.00& $ 7.6728\times 10^{-9} $&4.00& 3751\\
   			\hline
   			\multirow{4}{2cm}{BDF2}
   			&T/100 &$ 1.9126\times 10^{-3} $&     & $ 3.0096\times 10^{-2} $&    & 389\\
   			&T/200 &$ 8.4141\times 10^{-4} $& 1.18& $ 1.1454\times 10^{-2} $&1.39& 708\\
   			&T/400 &$ 2.3010\times 10^{-4} $& 1.87& $ 2.9890\times 10^{-3} $&1.94& 1230\\
   			&T/800 &$ 5.8409\times 10^{-5} $& 1.98& $ 7.5350\times 10^{-3} $&1.99& 2463\\
   			\hline
   			\multicolumn{7}{|c|}{$ tol_{rel,nonlinear}=10^{-8} $, $ tol_{rel,linear}=10^{-1} $}\\
   			\hline	
   			&$ \Delta t $ &  $ E_{L_2}( \rho )  $  & order  & $ E_{L_2}( u )  $  & order & CPU time(s)\\
   			\hline
   			\multirow{4}{2cm}{ESDIRK2}
   			&T/100 &$ 4.4088\times 10^{-4} $&     & $ 5.7335\times 10^{-3} $&     & 546\\
   			&T/200 &$ 1.1278\times 10^{-4} $& 1.97& $ 1.4549\times 10^{-3} $&1.98 & 930\\
   			&T/400 &$ 2.8267\times 10^{-5} $& 2.00& $ 3.6420\times 10^{-4} $&2.00 & 1601\\
   			&T/800 &$ 7.0691\times 10^{-6} $& 2.00& $ 9.1041\times 10^{-5} $&2.00 & 3003\\
   			\hline
   			\multirow{4}{2cm}{ESDIRK3}
   			&T/100 &$ 1.5165\times 10^{-4} $&     & $ 1.6911\times 10^{-3} $&    & 891\\
   			&T/200 &$ 2.3021\times 10^{-5} $& 2.72& $ 2.5399\times 10^{-4} $&2.74& 1373\\
   			&T/400 &$ 3.0913\times 10^{-6} $& 2.90& $ 3.3164\times 10^{-5} $&2.94& 2138\\
   			&T/800 &$ 3.9326\times 10^{-7} $& 2.97& $ 4.1867\times 10^{-5} $&2.99& 3867\\
   			\hline
   			\multirow{4}{2cm}{ESDIRK4}
   			&T/100 &$ 3.4502\times 10^{-6} $&     & $ 3.1249\times 10^{-5} $&    & 971\\
   			&T/200 &$ 2.1706\times 10^{-7} $& 3.99& $ 1.9593\times 10^{-6} $&4.00& 1550\\
   			&T/400 &$ 1.3584\times 10^{-8} $& 4.00& $ 1.2270\times 10^{-7} $&4.00& 2543\\
   			&T/800 &$ 8.4976\times 10^{-10} $& 4.00& $ 7.6747\times 10^{-9} $&4.00& 4370\\
   			\hline
   			\multirow{4}{2cm}{BDF2}
   			&T/100 &$ 1.9126\times 10^{-3} $&     & $ 3.0096\times 10^{-2} $&    & 455\\
   			&T/200 &$ 8.4141\times 10^{-4} $& 1.18& $ 1.1454\times 10^{-2} $&1.39& 786\\
   			&T/400 &$ 2.3010\times 10^{-4} $& 1.87& $ 2.9890\times 10^{-3} $&1.94& 1393\\
   			&T/800 &$ 5.8409\times 10^{-5} $& 1.98& $ 7.5350\times 10^{-3} $&1.99& 2651\\
   			\hline	
   		\end{longtable}
   	\end{center}	
    
    \clearpage
    \begin{table}
	\centering
	\caption{ The effect of $ tol_{rel,linear} $ on computational cost of ESDIRK.}
	\label{fullyimplicit_linear}
	\begin{tabular}{|c|c|r|r|r|r|}
		\hline
		& &  \multicolumn{4}{|c|}{CPU time (second)}\\	
		\hline
		&$tol_{rel,linear}$ & $ 10^{-1}$ & $ 10^{-2} $ & $ 10^{-4} $ & $ 10^{-6}$ \\
		\hline
		\multirow{4}{2cm}{ESDIRK2}
		&T/100 &464 & 503 & 693 &1041 \\
		&T/200 &814 & 856 & 1109&1551 \\
		&T/400 &1503& 1563& 1909&2501 \\
		&T/800 &2672& 2875& 3334&4203 \\
		\hline
		\multirow{4}{2cm}{ESDIRK3}
		&T/100 &692 & 686 & 1036&1041 \\
		&T/200 &1138& 1128& 1584&2419 \\
		&T/400 &1877& 1865& 2580&3695 \\
		&T/800 &3369& 3526& 4430&5908 \\
		\hline
		\multirow{4}{2cm}{ESDIRK4}
		&T/100 &780 & 875 & 1326&2095 \\
		&T/200 &1310 & 1410 & 2133&3004 \\
		&T/400 &2156& 2439& 3277&4844 \\
		&T/800 &3751& 4136& 5527&7636 \\
		\hline
	\end{tabular}
    \end{table}

    \subsubsection{Effect of GMRES convergence criterion on ROW}	
    	A comparison of different convergence criteria of the restarted GMRES solver is conducted to study its impact on the accuracy and efficiency of the ROW time integrators. It is observed that when the convergence criterion is not tight, such as $ tol_{rel,linear} = 10^{-2}$ and $10^{-4}$, the ROW methods cannot preserve the nominal order of accuracy. This is not a surprise considering that the residual convergence is directly related to the accuracy of the solution in ROW. When $ tol_{rel,linear}$ is tight, such as $ tol_{rel,linear} = 10^{-6}$ and $10^{-8}$, all ROW methods can preserve the nominal order of accuracy, excepted that ROW4 shows order reduction when the time step is refined from $T/400$ to $T/800$. We also notice that when $ tol_{rel,linear}$ is refined from $ 10^{-6} $ to $ 10^{-8} $, no significant differences in errors are observed. Another observation is that when $ tol_{rel,linear}$ is overrefined, the CPU time of the simulation is noticeably increased. In general, we do not recommend machine zero convergence criterion for $ tol_{rel,linear}$. A relatively tight value such as $ 10^{-6} $ is sufficient to preserve the accuracy of high-order Rosenbrock methods for this problem.

	\clearpage
    \begin{center}
	\begin{longtable}{|c|c|c|c|c|c|c|}
		\caption{The convergence study for ROW methods with different $ tol_{rel,linear} $.}	\label{linearly_implicit}\\		
		
		\endfirsthead
		
		\hline
		\multicolumn{7}{|l|}{Continuation of Table~\ref{linearly_implicit},}\\
		\hline
		\endhead
		
		\multicolumn{7}{|r|}{To be continued on next page.}\\
		\hline
		\endfoot
		
		\multicolumn{7}{| r |}{End of Table~\ref{linearly_implicit}.}\\
		\hline
		\endlastfoot 
		
		\hline
		\multicolumn{7}{|c|}{$ tol_{rel,linear}=10^{-2} $}\\
		\hline	
		&$ \Delta t $ &  $ E_{L_2}( \rho )  $  & order  & $ E_{L_2}( u )  $  & order & CPU time(s)\\
		\hline
		\multirow{4}{2cm}{ROW2}
		&T/100 & Diverged &   --  & -- &   --  &  \\
		&T/200 &$ 2.3615\times 10^{-4} $&     & $ 3.3198\times 10^{-3} $&     & 537\\
		&T/400 &$ 9.9393\times 10^{-5} $& 1.25& $ 1.3753\times 10^{-3} $&1.27 & 1052\\
		&T/800 &$ 5.5248\times 10^{-5} $& 0.85& $ 7.4627\times 10^{-4} $&0.88 & 2073\\
		\hline
		\multirow{4}{2cm}{ROW3}
		&T/100 &$ 2.6132\times 10^{-4} $&     & $ 2.6888\times 10^{-3} $&    & 300\\
		&T/200 &$ 8.8493\times 10^{-5} $& 1.56& $ 8.3870\times 10^{-4} $&1.68& 564\\
		&T/400 &$ 3.3446\times 10^{-5} $& 1.40& $ 4.0554\times 10^{-4} $&1.05& 1124\\
		&T/800 &$ 9.1969\times 10^{-6} $& 1.86& $ 1.1447\times 10^{-4} $&1.82& 2191\\
		\hline
		\multirow{4}{2cm}{ROW4}
		&T/100 &$ 3.6161\times 10^{-5} $&     & $ 4.1926\times 10^{-4} $&     & 344\\
		&T/200 &$ 9.0325\times 10^{-6} $& 2.00& $ 8.3350\times 10^{-5} $&2.33 & 638\\
		&T/400 &$ 2.7913\times 10^{-6} $& 1.69& $ 2.7282\times 10^{-5} $&1.61 & 1194\\
		&T/800 &$ 3.9397\times 10^{-6} $&-0.50& $ 5.1692\times 10^{-5} $&-0.92& 2282\\
		\hline
		
		\multicolumn{7}{|c|}{$ tol_{rel,linear}=10^{-4} $}\\
		\hline	
		&$ \Delta t $ &  $ E_{L_2}( \rho )  $  & order  & $ E_{L_2}( u )  $  & order & CPU time(s)\\
		\hline
		\multirow{4}{2cm}{ROW2}
		&T/100 & Diverged &   --  & -- &  --   &  \\
		&T/200 &$ 8.6277\times 10^{-5} $&     & $ 1.1163\times 10^{-3} $&     & 641\\
		&T/400 &$ 2.1800\times 10^{-5} $& 1.98& $ 2.8024\times 10^{-3} $&1.99 & 1189\\
		&T/800 &$ 5.4708\times 10^{-6} $& 1.99& $ 7.1383\times 10^{-5} $&1.97 & 2255\\
		\hline
		\multirow{4}{2cm}{ROW3}
		&T/100 &$ 1.5467\times 10^{-4} $&     & $ 1.7185\times 10^{-3} $&    & 449\\
		&T/200 &$ 2.3151\times 10^{-5} $& 2.74& $ 2.5539\times 10^{-4} $&2.75& 779\\
		&T/400 &$ 3.1137\times 10^{-6} $& 2.89& $ 3.3395\times 10^{-5} $&2.94& 1382\\
		&T/800 &$ 4.0729\times 10^{-7} $& 2.93& $ 4.2406\times 10^{-6} $&2.98& 2565\\
		\hline
		\multirow{4}{2cm}{ROW4}
		&T/100 &$ 4.0016\times 10^{-6} $&     & $ 3.8660\times 10^{-5} $&     & 487\\
		&T/200 &$ 3.5565\times 10^{-7} $& 3.49& $ 3.4725\times 10^{-6} $& 3.47& 846\\
		&T/400 &$ 2.7298\times 10^{-8} $& 3.70& $ 1.6676\times 10^{-7} $& 4.38& 1563\\
		&T/800 &$ 1.8642\times 10^{-8} $& 0.55& $ 2.2003\times 10^{-7} $&-0.40& 2769\\
		\hline
		\pagebreak
		\hline
		\multicolumn{7}{|c|}{$ tol_{rel,linear}=10^{-6} $}\\
		\hline	
		&$ \Delta t $ &  $ E_{L_2}( \rho )  $  & order  & $ E_{L_2}( u )  $  & order & CPU time(s)\\
		\hline
		\multirow{4}{2cm}{ROW2}
		&T/100 & Diverged &   --  & -- &  --   &  \\
		&T/200 &$ 8.6820\times 10^{-5} $&     & $ 1.1229\times 10^{-3} $&     & 767\\
		&T/400 &$ 2.1813\times 10^{-5} $& 1.99& $ 2.8148\times 10^{-4} $&2.00 & 1377\\
		&T/800 &$ 5.4628\times 10^{-6} $& 2.00& $ 7.0399\times 10^{-5} $&2.00 & 2500\\
		\hline
		\multirow{4}{2cm}{ROW3}
		&T/100 &$ 1.5474\times 10^{-4} $&     & $ 1.7185\times 10^{-3} $&    & 690\\
		&T/200 &$ 2.3165\times 10^{-5} $& 2.74& $ 2.5534\times 10^{-4} $&2.75& 1036\\
		&T/400 &$ 3.1047\times 10^{-6} $& 2.90& $ 3.3251\times 10^{-5} $&2.94& 1712\\
		&T/800 &$ 4.0081\times 10^{-7} $& 2.95& $ 4.1768\times 10^{-6} $&2.99& 3004\\
		\hline
		\multirow{4}{2cm}{ROW4}
		&T/100 &$ 3.9102\times 10^{-6} $&     & $ 3.6770\times 10^{-5} $&    & 682\\
		&T/200 &$ 2.5386\times 10^{-7} $& 3.95& $ 2.3433\times 10^{-6} $&3.97& 1091\\
		&T/400 &$ 2.5819\times 10^{-8} $& 3.30& $ 1.3773\times 10^{-7} $&4.09& 1849\\
		&T/800 &$ 7.5834\times 10^{-9} $& 1.78& $ 1.7492\times 10^{-8} $&2.98& 3184\\
		\hline
		\multicolumn{7}{|c|}{ $ tol_{rel,linear}=10^{-8} $}\\
		\hline	
		&$ \Delta t $ &  $ E_{L_2}( \rho )  $  & order  & $ E_{L_2}( u )  $  & order & CPU time(s)\\
		\hline
		\multirow{4}{2cm}{ROW2}
		&T/100 & Diverged & --    & -- &  --   &  \\
		&T/200 &$ 8.6824\times 10^{-5} $&     & $ 1.1229\times 10^{-3} $&     & 906\\
		&T/400 &$ 2.1810\times 10^{-5} $& 1.99& $ 2.8147\times 10^{-4} $&2.00 & 1557\\
		&T/800 &$ 5.4619\times 10^{-6} $& 2.00& $ 7.0398\times 10^{-5} $&2.00 & 2722\\
		\hline
		\multirow{4}{2cm}{ROW3}
		&T/100 &$ 1.5476\times 10^{-4} $&     & $ 1.7186\times 10^{-3} $&    & 862\\
		&T/200 &$ 2.3158\times 10^{-5} $& 2.74& $ 2.5535\times 10^{-4} $&2.75& 1589\\
		&T/400 &$ 3.1015\times 10^{-6} $& 2.90& $ 3.3242\times 10^{-5} $&2.94& 2149\\
		&T/800 &$ 4.0093\times 10^{-7} $& 2.95& $ 4.2024\times 10^{-6} $&2.98& 3481\\
		\hline
		\multirow{4}{2cm}{ROW4}
		&T/100 &$ 3.9055\times 10^{-6} $&     & $ 3.6728\times 10^{-5} $&    & 1186\\
		&T/200 &$ 2.5299\times 10^{-7} $& 3.96& $ 2.3382\times 10^{-6} $&3.97& 1381\\
		&T/400 &$ 2.6107\times 10^{-8} $& 3.28& $ 1.4352\times 10^{-7} $&4.03& 2237\\
		&T/800 &$ 7.5988\times 10^{-9} $& 1.78& $ 1.7442\times 10^{-8} $&3.04& 3720\\
		\hline
	\end{longtable}
    \end{center}
	
	\subsubsection{Comparison of different time integrators}
\begin{figure}		
	\centering
	\begin{tabular}{cc}
		\includegraphics[height=6.5cm]{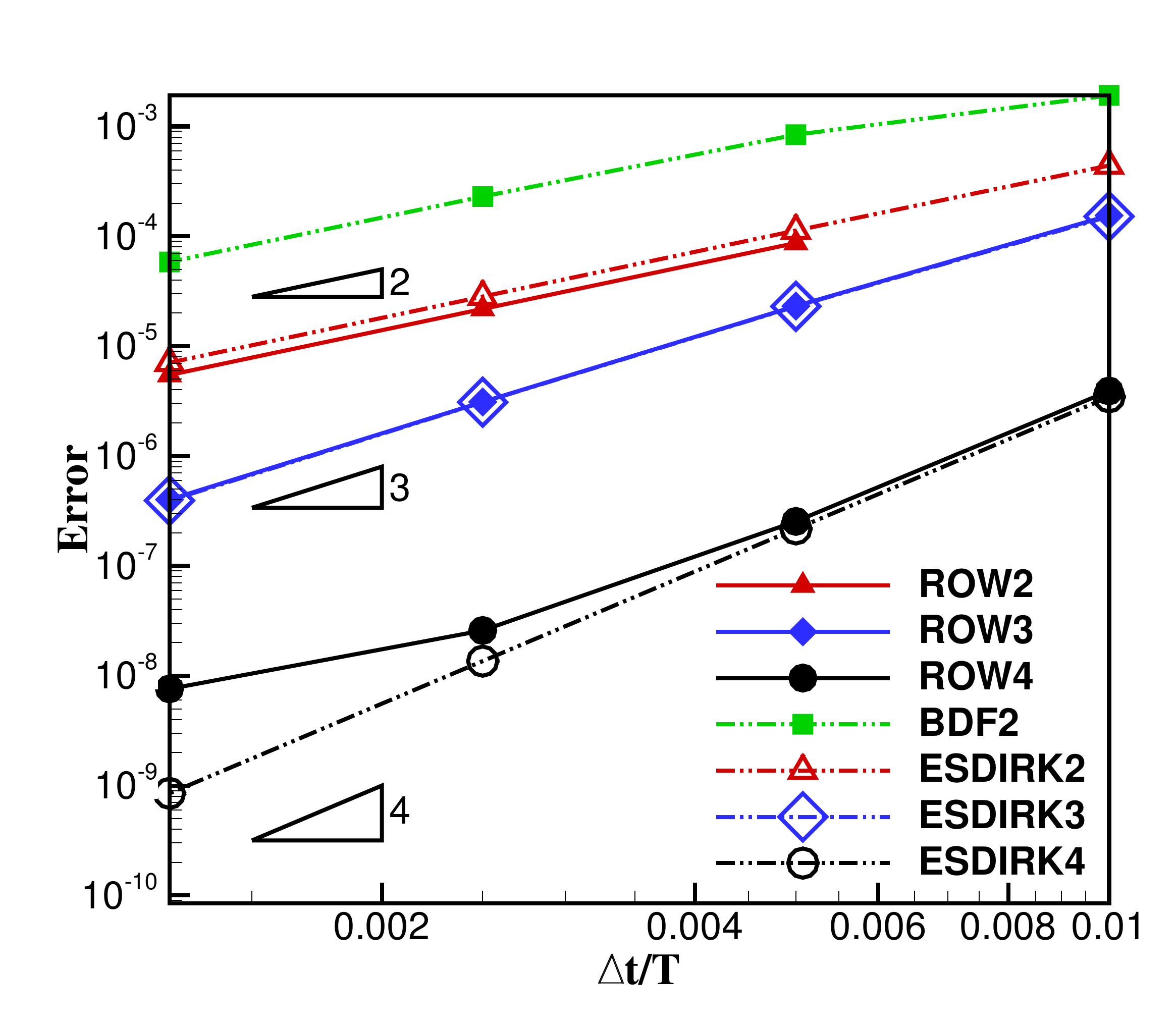}\\
		(a) \\
		\includegraphics[width=12cm]{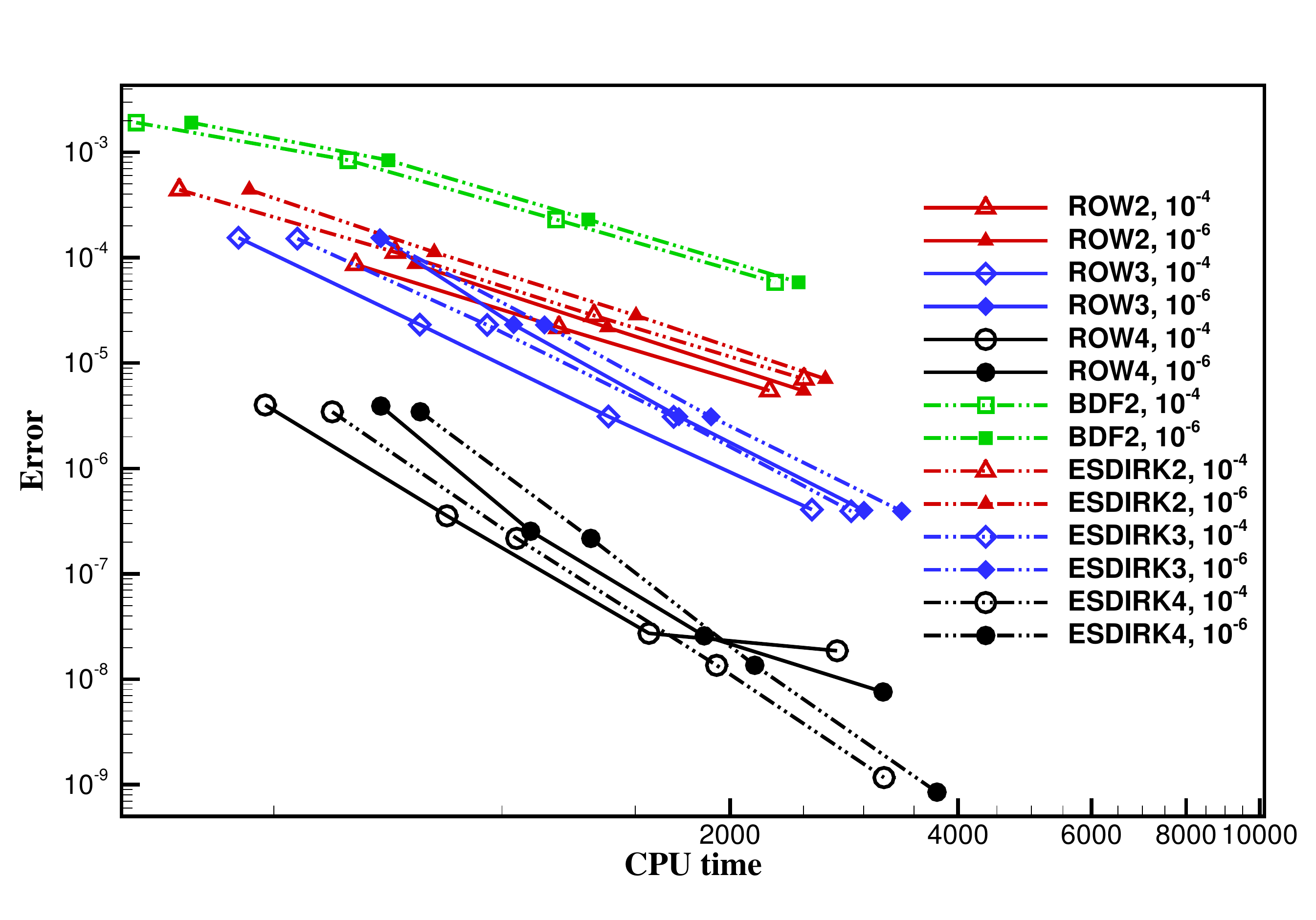}\\
	    (b) \\
	\end{tabular}
	\caption{(a) The convergence study of different time integrators and (b) efficiency study of different time integrators for the vortex propagation problem.}
	\label{vortex_convergence_all}	
	\end{figure}

	The convergence study of  ESDIRK and BDF2 and linearly implicit ROW are summarized in Figure~\ref{vortex_convergence_all}. For ESDIRK and BDF2, $ tol_{rel,nonlinear}^{ESDIRK,BDF2} = 10^{-6} $ and $ tol_{rel,linear}^{ESDIRK,BDF2} = 10^{-1} $. For Rosenbrock methods, $ tol_{rel,linear}^{ROW} =10^{-6}$.  
	
	Figure~\ref{vortex_convergence_all}(b) presents the CPU time versus errors of different time integrators with different convergence criteria. As illustrated in Figure~\ref{vortex_convergence_all}(b), all multistage methods are significantly more efficient than BDF2. We notice that ROW2 and ESDIRK2 intersect with ROW3 and ESDIRK3. However, as the error threshold is decreased, higher-order methods will be more efficient.

	Generally speaking, when $ tol_{rel,nonlinear}^{ESDIRK} $ is the same as $ tol_{rel,linear}^{ROW} $, ROW methods are more efficient than ESDIRK methods. However, ROW methods cannot preserve the nominal order of accuracy when the convergence criterion is not tight, and even  suffer from severe order reduction~\cite{rang2015improved}. 
	Instead, $ tol_{rel,nonlinear} = 10^{-4} $ can make ESDIRK methods preserve the nominal order of accuracy (0.46 order reduction at most). It is observed that when $ tol_{rel,linear}^{ROW} = 10^{-6} $ and $ tol_{rel,nonlinear}^{ESDIRK} = 10^{-4} $ are employed, ESDIRK methods are more efficient than ROW methods. 
	This indicates that ESDIRK methods tend to be over-solved more easily than ROW methods if the nonlinear convergence criterion $tol_{rel,nonlinear}^{ESDIRK}$ is not set up judiciously.	
	We have also noticed that ESDIRK can be more robust than ROW. As documented in Table~\ref{fullyimplicit_nonlinear} and Table~\ref{linearly_implicit}, when the tolerance criteria are set to $10^{-2}$, ESDIRK2 can show optimal convergence rate, while ROW cannot preserve it; when  $ \Delta t = T/100 $, ROW2 even fails to converge.

	\subsection{Laminar flow over a circular cylinder}
\begin{figure}		
		\centering
		\includegraphics[width=12cm]{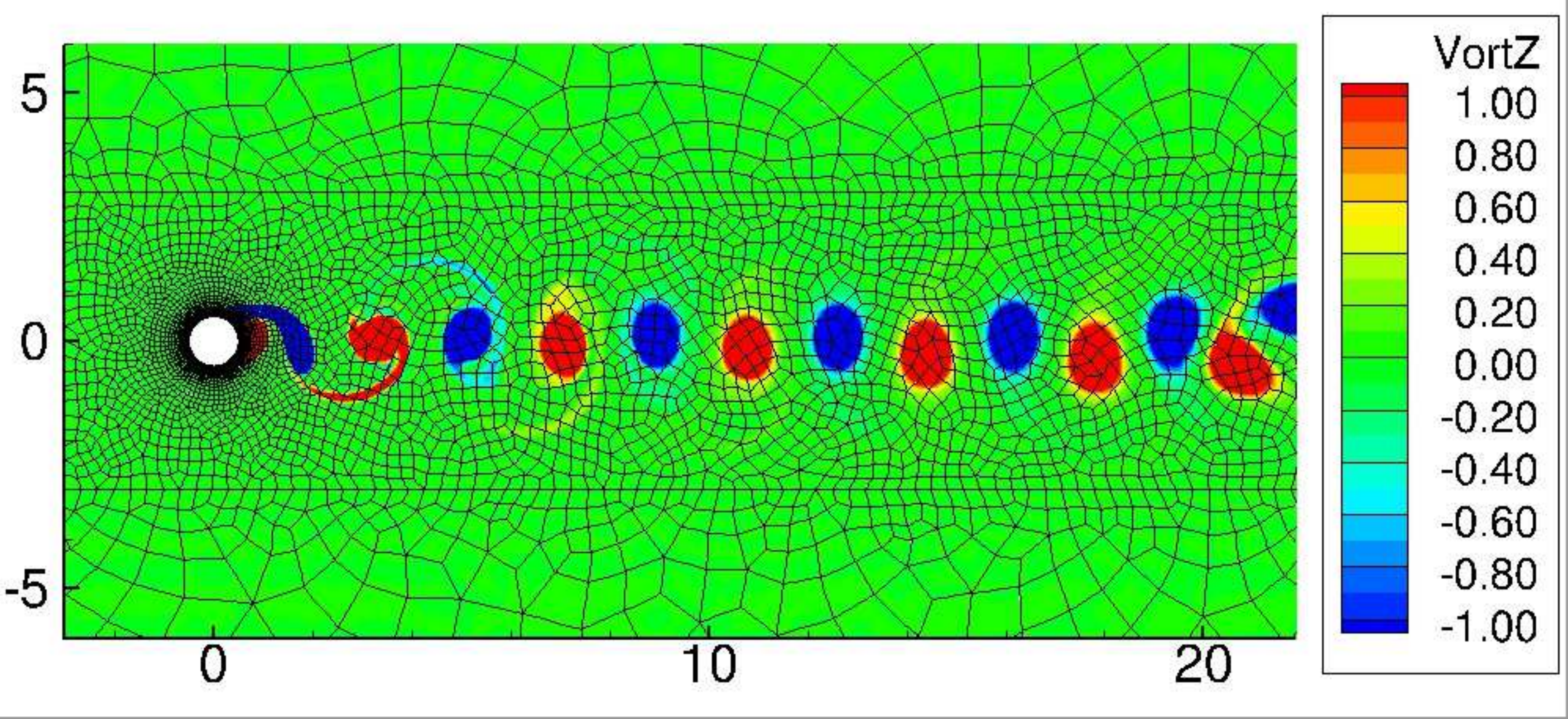}

		\caption{The mesh and an instance of the wake of vortex shedding for the laminar flow over a circular cylinder at $Re=1200$.}
		\label{cylinder-mesh}
	\end{figure}

	In this section, we employ the laminar flow over the circular cylinder as an example to study the performance of different time integrators. This case has been tested in various literature~\cite{bijl2002implicit,blom2016comparison}. The flow conditions are set as $ Ma=0.1 $ and $ Re=1200 $, where $Ma$ stands for Mach number, and $Re$ stands for Reynolds number. The diameter of the cylinder is set as $ 1 $, and computational domain is $ [-100,200]\times[-100,100] $. The mesh in the near wall region and the wake region, and an instance of the vortex shedding are presented in Figure~\ref{cylinder-mesh}. There are 5690 elements in the mesh. The height of the first layer of the mesh is roughly $ 0.0033 $.
	The $ P^3 $ FR method is employed for the spatial discretization. 	
	$ \Delta \tau_{init} = 0.01 $ and $ \Delta \tau_{max}={1.0} $ are used for all simulations in this section. A moderate $ \Delta \tau_{max} $ is to prevent that GMRES solver can not even converge by one magnitude when $ \Delta \tau = \Delta \tau_{max} $.  $ tol_{rel,linear}^{ROW} = tol_{rel,nonlinear}^{ESDIRK, BDF2} = 10^{-6} $, and $ tol_{rel,linear}^{ESDIRK, BDF2} = 10^{-1} $. As aforementioned, for ESDIRK and BDF, if the linear solver fails to drive the residual to drop by one magnitude, the current iteration in the  pseudo-transient continuation procedure will decrease $ \Delta \tau_{max} $ by half and restart.
	
\begin{figure}		
		\centering
		\begin{tabular}{cc}
			\hspace{-0.6cm}
			\includegraphics[width=7cm]{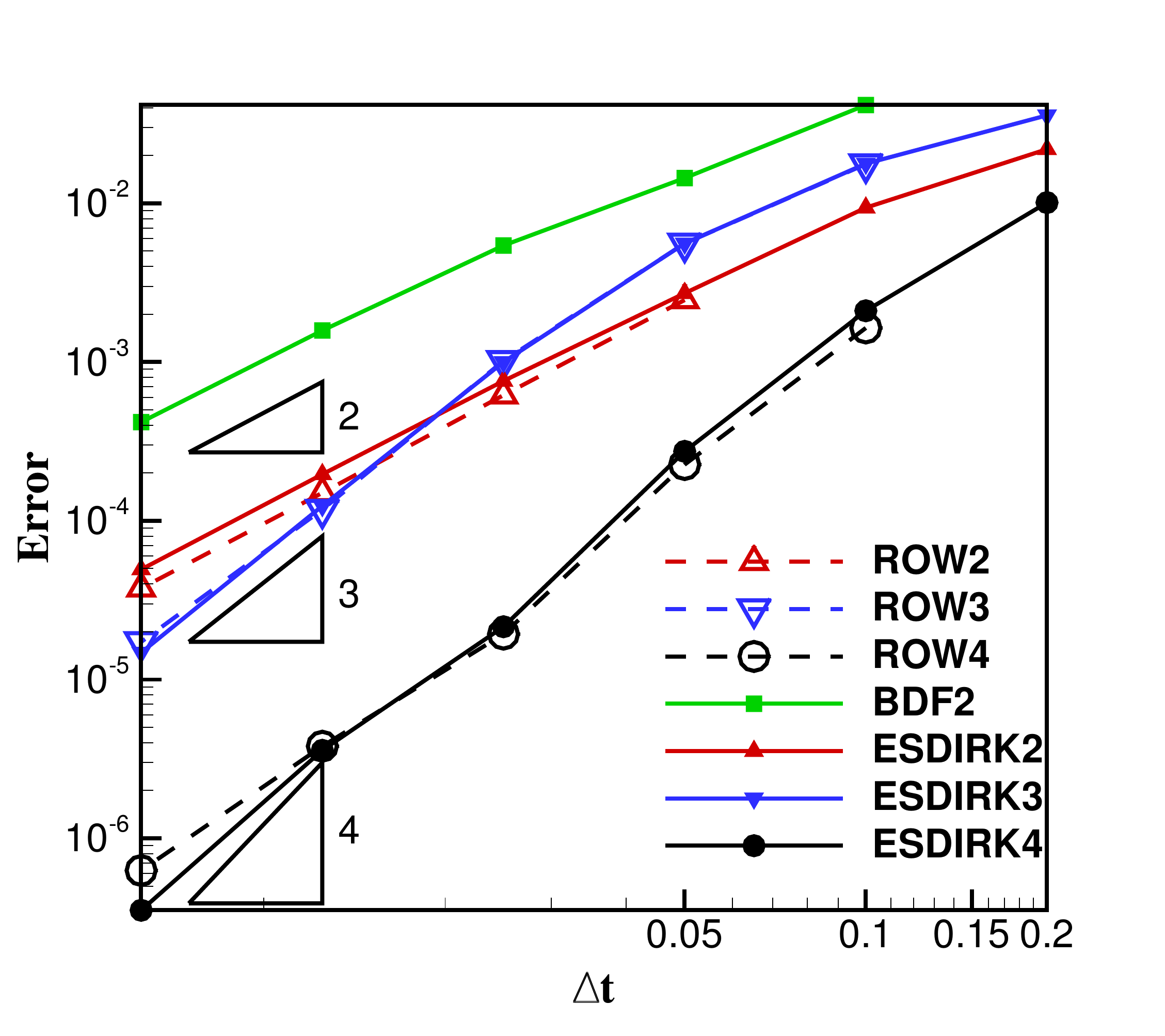} &
			\includegraphics[width=7cm]{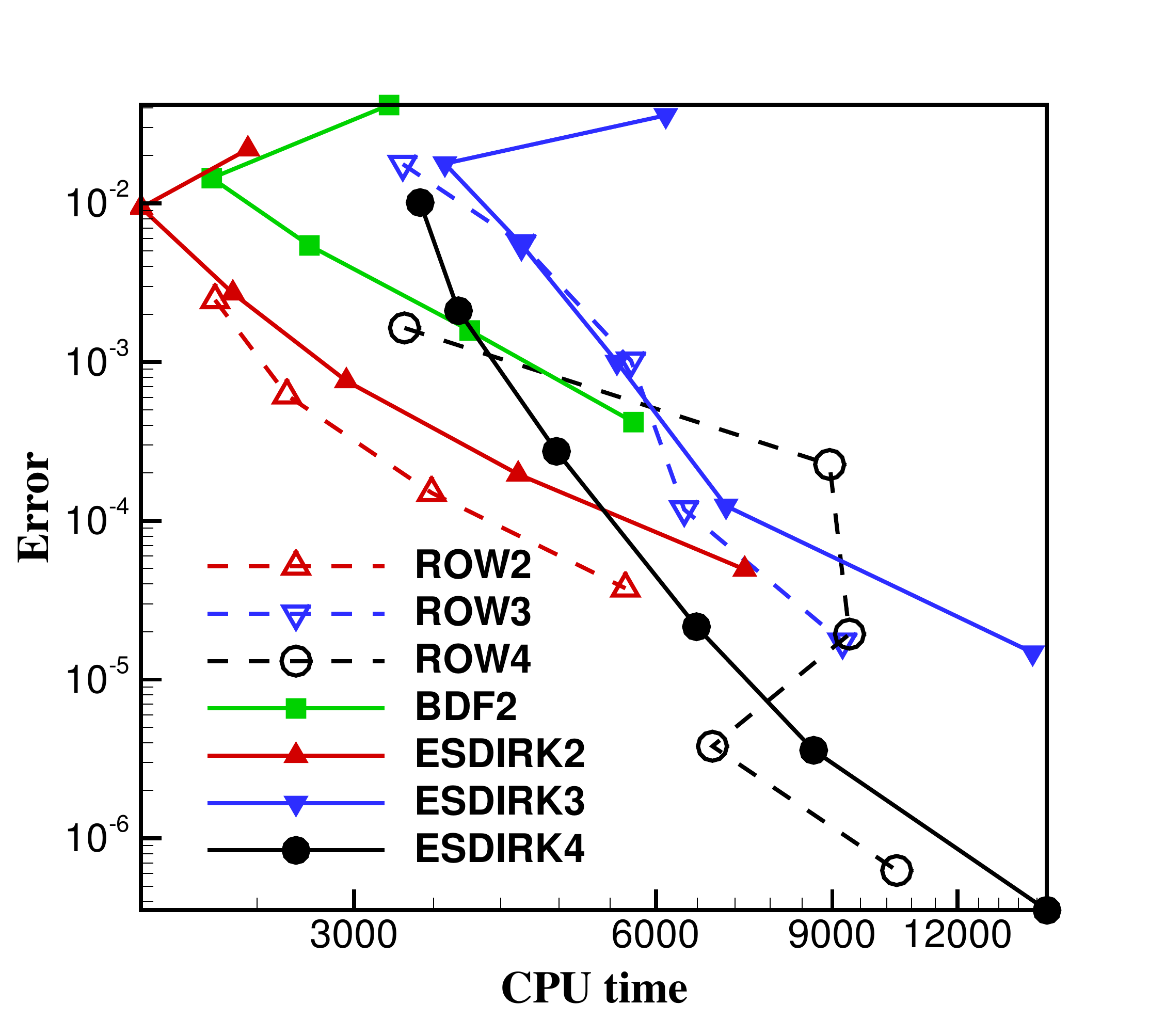}\\
			(a)& (b) \\
		\end{tabular}
		\caption{The convergence and efficiency study for the laminar flow over the circular cylinder. (a) Error vs. time step size $ \Delta t $ and (b) error vs. CPU time.}
		\label{flow_over_cylinder}
	\end{figure}

	The flow is initialized with the steady solution at $ Re=40 $. And then we use ESDIRK4 to run this simulation untill $t=180$ with $ \Delta t = 0.001 $ to obtain the initial conditions for the convergence and efficiency study. For the convergence and efficiency study, we run all simulations for ten seconds along the physical time. The time step size is refined from $ 0.2 $ to $ 0.00625 $.
	We use the numerical results of explicit SSPRK3 with a small time step $ \Delta t = 5\times10^{-6} $  as the reference value. The drag  coefficient $ C_d $ is used for the error estimation. The error is calculated as 
	\begin{equation}
	Error(C_d)= \sqrt{\frac{\sum_{n=1}^{N}(C_{d,n}-C_{d,ref,n})^2}{N}},
	\end{equation}
	where $C_{d,ref,n}$ is the reference value from SSPRK3, and $ N $ is the number of time steps. The results from convergence and efficiency study are presented in Figure~\ref{flow_over_cylinder}(a) and 
	Figure~\ref{flow_over_cylinder}(b), respectively. As the time step size is refined, all second- and third-order methods will converge at the nominal convergence rate. For both ROW4 and ESDIRK4, we have observed order reduction. The order reduction of ROW4 is more severe than ESDIRK4. In terms of CPU time, when $ \Delta t = 0.2 $, unexpected computational cost is observed for ESDIRK and BDF methods. Many iterations in the pseudo-transient continuation are rejected due to the poor performance of the element-Jacobi preconditioner.  However, all ROW methods fail when $ \Delta t = 0.2 $ and ROW2 even fails when $ \Delta t = 0.1 $. At a relatively large error level, such as $ 10^{-3} $, second-order methods take the least mount of time. However, to reach a lower error level, higher-order methods are more efficient. We have noticed that the abnormal increase in CPU time for ROW4 when $ \Delta t=0.05 $ and $ \Delta t=0.025 $. This is due to the fact that the residual of the restarted GMRES solver with an element-Jacobi preconditioner sometimes cannot converge to the designated $ tol_{rel,linear} $ when the maximum number of iteration is reached. From this study, we find that the performance of schemes from the ESDIRK family is more consistent than that of ROW methods when simulating unsteady flows over walls; and when the requirement of the error threshold is stringent, such as $10^{-6}$ in this case, ROW can be more efficient than ESDIRK.    
	
	\subsection{Taylor-Green vortex}
\begin{figure}		
	\centering
	\includegraphics[height=8cm]{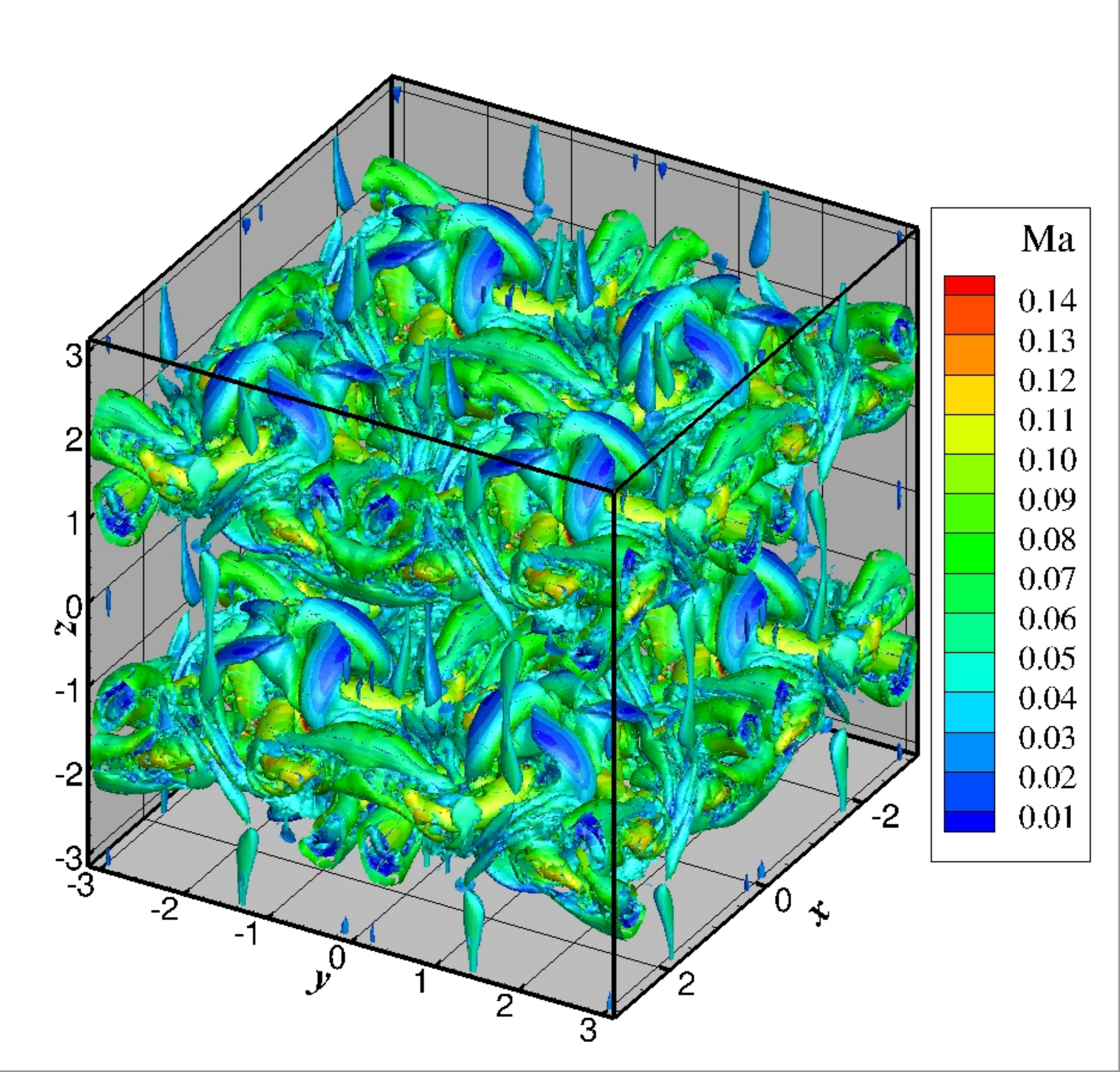}
	\caption{The isosurface of the Q criterion with $ Q_{criterion} = 1 $ colored by $ Ma $ at $ t= 8 t_c $ for Taylor-Green vortex evolution.}
	\label{q_criterion}
	\end{figure}

	The Taylor-Green vortex is a benchmark to test the accuracy and performance of high-order methods on the direct numerical simulation of a 3D periodic and transitional flow defined by
	initial conditions~\cite{liu2016comparative}
	\begin{equation}	
	\begin{cases}
	u =\ V_0\sin(x/L)\cos(y/L)\cos(z/L),\\
	v =-V_0 \cos(x/L)\sin(y/L)\cos(z/L),\\
	w =0,\\
	p=p_0+\frac{\rho_0V_0^2}{16}\left(\cos(2x/L)+\cos(2y/L)\right)\left(\cos(2z/L)+2\right).	
	\end{cases}
	\end{equation}
	The domain is $ \Omega = [-\pi L, \pi L]^3 $. 
	The Reynolds number of the flow is defined as $ Re=\frac{\rho_0V_0L}{\mu} $ and is equal to 1600. For this study, we consider the flow with weak compressibility and the perfect gas law holds, i.e., $ p = \rho R T$. The Prandtl number is $ Pr=\frac{\mu C_p}{\kappa}=0.71 $. We assume that the gas has zero bulk viscosity $ \mu_v = 0 $. The Mach number $ Ma=\frac{V_0}{c_0}=0.1 $, where $ c_0 $ is the speed of sound corresponding to $ p_0 $. The characteristic convection time is defined as $ t_c = \frac{L}{V_0} $. The maximum dissipation occurs at $ t\approx 8 t_c $. An uniform $ 64^3 $ mesh is employed and the $ P^3 $ FR methods is used for simulation. We conduct all simulation until $ t=10t_c $. Figure~\ref{q_criterion} presents the isosurface of the Q criterion with $ Q_{criterion} = 1 $ colored by $ Ma $ at $ t= 8 t_c $.
	
	We employ the error of kinetic energy dissipation rate for the accuracy and efficiency study. The kinetic energy dissipation rate  of compressible flows is the summation of three contributions as $	\epsilon = \epsilon_1+\epsilon_2+\epsilon_3 $:
	\begin{equation}
	\epsilon_1 = 2\frac{\mu}{\rho_0}\frac{1}{V}\int_\Omega \boldsymbol{S}^d : \boldsymbol{S}^dV, 
	\end{equation}
	where $ \boldsymbol{S}^d $ is the deviatoric part of the strain rate tensor, and $V$ is the volume of the domain $\Omega$,
	\begin{equation}
	\epsilon_2 = \frac{\mu_v}{\rho_0}\frac{1}{V}\int_\Omega (\nabla\cdot \boldsymbol{v})^2dV,
	\end{equation}
	where $ \mu_v =0 $, and
	\begin{equation}
	\epsilon_3 = -\frac{1}{\rho_0}\frac{1}{V}\int_\Omega p\nabla \cdot \boldsymbol{v} dV.
	\end{equation}
	The numerical results of SSPRK3 method with $ \Delta t = 2\times10^{-4} t_c$ is adopted as the reference data for error evaluations. We define the error of the kinetic energy dissipation rate as 
	\begin{equation}
	Error(\epsilon)= \sqrt{\frac{\sum_{n=1}^{N}(\epsilon_n-\epsilon_{ref,n})^2}{N}},
	\end{equation}	
	where $\epsilon_{ref,n}$ is the reference value from SSPRK3, and $ N $ is the number of time steps.
	
	As discovered in previous sections, the convergence criteria have significant effect on the efficiency of ROW, ESDIRK and BDF2. In this section, we only consider $ tol_{rel,linear} $ of ROW is the same as $ tol_{rel, nolinear} $ of ESDIRK and BDF2. Herein, $ tol_{rel,linear}^{ROW} = tol_{rel,nonlinear}^{ESDIRK, BDF2} = 10^{-6} $. For the inexactly linear-solving part of ESDIRK and BDF2, we employ $ tol_{rel,linear}^{ESDIRK, BDF2} = 10^{-1} $ to save computational cost.
	
\begin{figure}		
		\centering
		\begin{tabular}{cc}
		\hspace{-0.6cm}
			\includegraphics[width=7cm]{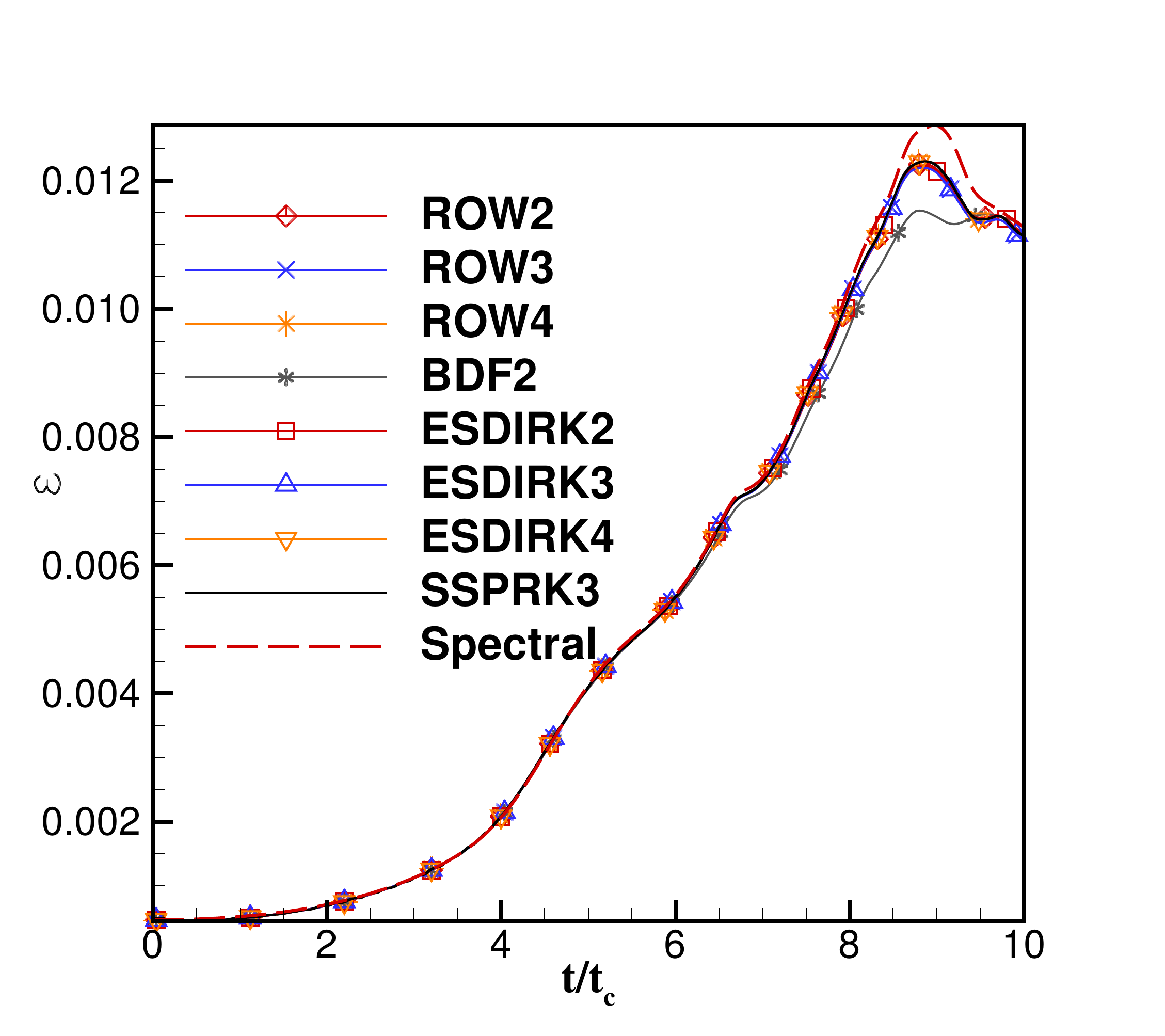} &
			\includegraphics[width=7cm]{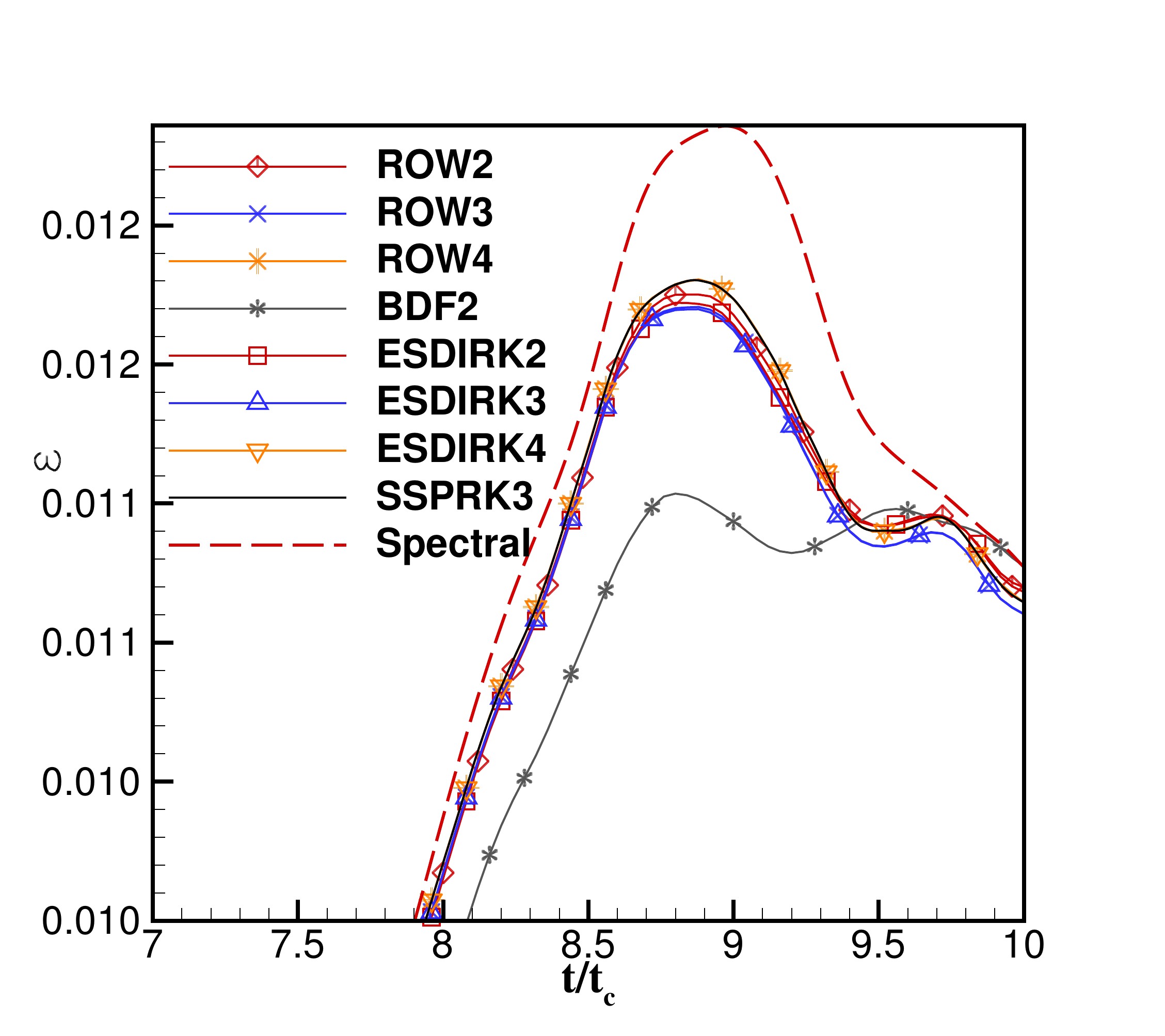}\\
			(a)& (b) \\
		\end{tabular}
		\caption{The kinetic energy dissipation rate history of Taylor-Green vortex decaying. (a) A full view when $ t/t_c \in [0,10] $; (b) a close-up view when $ t/t_c \in [7,10] $.}
		\label{close-up-view-epsilon}
	\end{figure}

\begin{figure}		
	\centering
	\begin{tabular}{cc}
	\hspace{-0.6cm}
	\includegraphics[width=7cm]{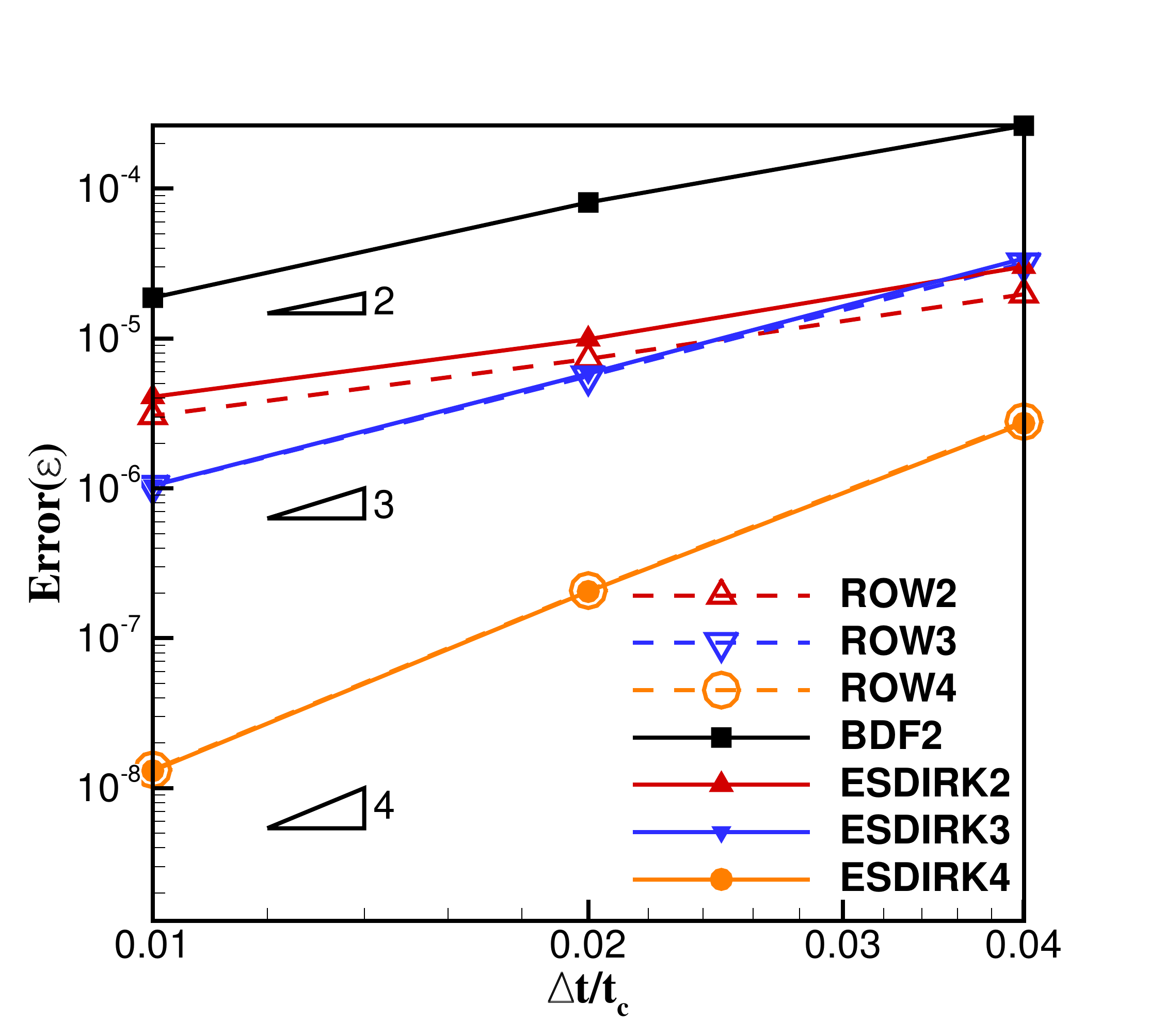} &
	\includegraphics[width=7cm]{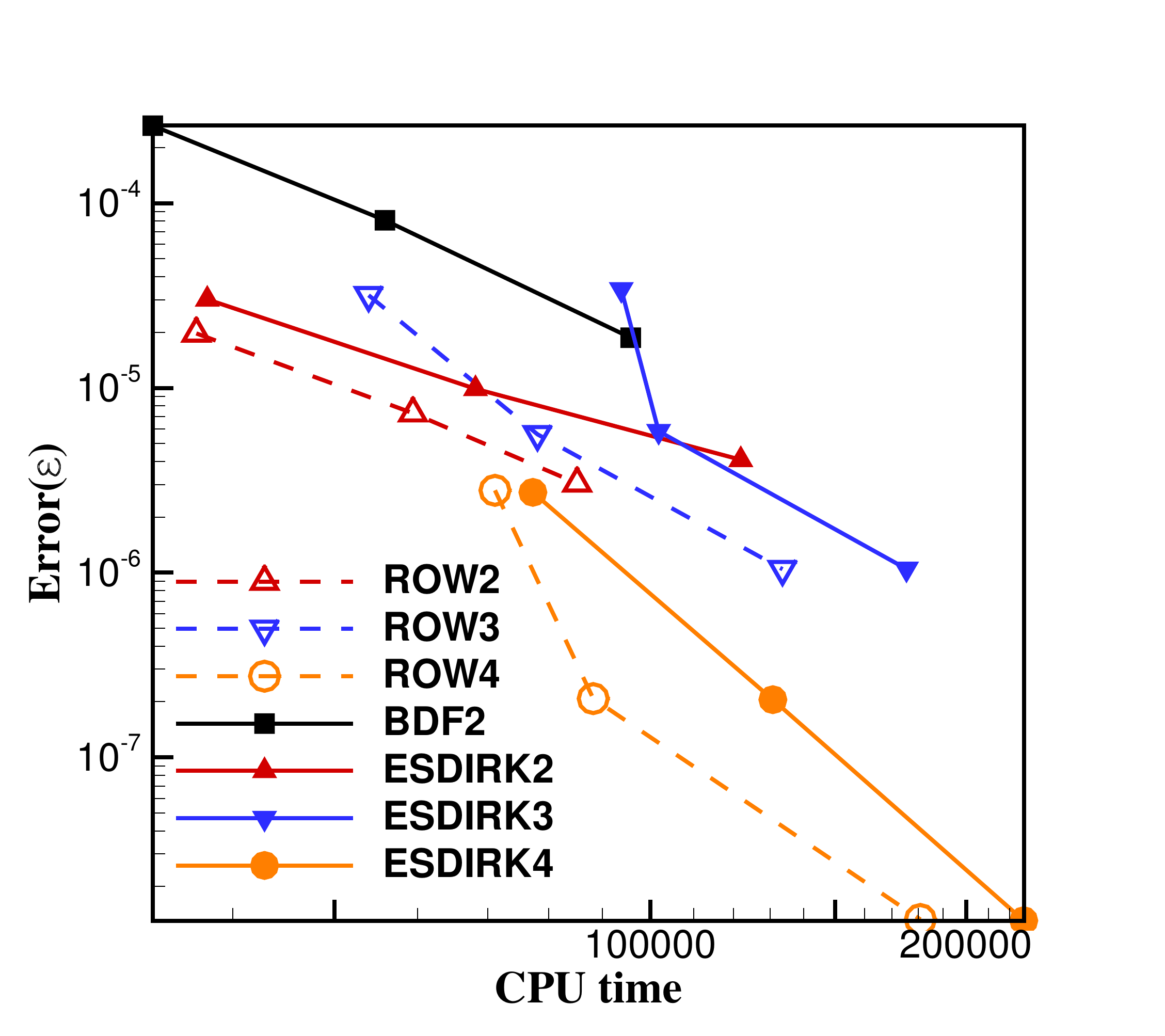}\\
	(a)& (b) \\
	\end{tabular}
	\caption{The convergence and efficiency study for Taylor-Green vortex evolution. (a) Error vs. time step size $ \Delta t/t_c $ and (b) error vs. CPU time.}
	\label{taylor_green_time_refinement}
	\end{figure}

	The time step size $ \Delta t $ is refined from $ t_c/25 $ to $ t_c/100 $. The kinetic energy dissipation rate history when $ \Delta t = t_c/25 $ is presented in Figure~\ref{close-up-view-epsilon}(a). 
	A close-up view within $ t/t_c \in [8.5,10] $ is illustrated in Figure~\ref{close-up-view-epsilon}(b). The numerical results from the spectral method on a $ 512^3 $ mesh is also presented for reference~\cite{van2011comparison}. Our observation is that the results of ROW4 and ESDIRK4 almost coincide with that of SSPRK3. The result of BDF2 is the least accurate.  The convergence study is presented in  Figure~\ref{taylor_green_time_refinement}. Figure~\ref{taylor_green_time_refinement}(a) shows the error vs. time step size $ \Delta t/t_c $ and Figure~\ref{taylor_green_time_refinement}(b) shows the error vs. CPU time. From Figure~\ref{taylor_green_time_refinement}(a), the convergence features of ROW and ESDIRK are almost the same.  As is shown in Figure~\ref{taylor_green_time_refinement}(b), BDF2 is not efficient as expected. 
	When $ tol_{rel,linear}^{ROW} = tol_{rel,nonlinear}^{ESDIRK, BDF2} = 10^{-6} $, ROW methods are found to be more efficient than ESDIRK methods. From Figure~\ref{close-up-view-epsilon}, we find that for turbulent simulation, compared to the results from SSPRK3 with very small time steps, all time integrators with excessively large time step size will lead to numerical dissipation of the kinetic energy dissipation rate except ROW4 and ESDIRK4. This reveals that the dissipation due to the temporal disretization should also be taken into account for turbulent simulation. We refers interested readers to Refs.~\cite{vermeire2017behaviour,ZJW2018} for more discussions.

	\section{Conclusions}
	In this study, we compare the accuracy and efficiency of ROW and ESDIRK (from second order to fourth order) and BDF2 for unsteady flow simulation with high-order FR formulations. We find that the multistage time integrators, i.e. ROW and ESDIRK, are more computationally efficient than BDF2. When we compare ROW and ESDIRK, the efficiency of ROW and ESDIRK highly depends on the convergence criteria of solving nonlinear equations and linear equations.  For most of the case, when $ tol_{rel,linear}^{ROW} = tol_{rel,nonlinear}^{ESDIRK} $, ROW is more efficient. However, since the residual convergence is directly related to the accuracy of the solution with ROW, the tolerance $ tol_{rel,linear}^{ROW} $ needs to be tight for unsteady flow simulation. By contrast, the tolerance $ tol_{rel,nonlinear}^{ESDIRK} $ for pseudo-time iterations in ESDIRK needs not to be tight, and the tolerance $ tol_{rel,linear}^{ESDIRK} $ for the restarted GMRES linear solver can always be set to a large value, i.e. $10^{-1}$ in all cases tested here.  
	Therefore, when $tol_{rel,nonlinear}^{ESDIRK} $ is allowed to be larger than $ tol_{rel,linear}^{ROW}$, ESDIRK can be more efficient than ROW. Besides, ROW methods are easier to show order reduction than ESDIRK methods, making their performance less consistent than that of ESDIRK methods when flow problems become complicated, such as unsteady flows over walls. 
	We also find that the performance of the preconditioner can substantially affect the efficiency of ROW methods. Since the element-Jacobi preconditioner is used in this study, time steps for ROW cannot be very large in order to maintain good performance. Instead, ESDIRK methods generally allows a larger time step size for unsteady simulations than ROW methods.
	
	 

	\section*{Acknowledgment}
	The authors gratefully acknowledge the support of the Office of Naval Research through
	the award N00014-16-1-2735, and the faculty startup support from the department of
	mechanical engineering at the University of Maryland, Baltimore County (UMBC). The hardware used in the computational studies is part of the
	UMBC High Performance Computing Facility (HPCF).
	The facility is supported by the U.S. National Science Foundation
	through the MRI program
	(grant nos.~CNS-0821258, CNS-1228778, and OAC-1726023)
	and the SCREMS program (grant no.~DMS-0821311),
	with additional substantial support from UMBC.
	\clearpage

	\bibliographystyle{ieeetr}
	\bibliography{comparative_rosenbrock}

\begin{thebibliography}{10}

\bibitem{CockburnShu89}
{B. Cockburn and C.-W. Shu}, ``{TVB Runge-Kutta local projection discontinuous
  Galerkin finite element method for conservation laws II: general
  framework},'' {\em {Math. Comput.}}, vol.~52, pp.~411--435, 1989.

\bibitem{bassi1997high}
F.~Bassi and S.~Rebay, ``{A high-order accurate discontinuous finite element
  method for the numerical solution of the compressible Navier--Stokes
  equations},'' {\em Journal of computational physics}, vol.~131, no.~2,
  pp.~267--279, 1997.

\bibitem{NDG08}
J.~S. Hesthaven and T.~Warburton, {\em Nodal Discontinuous Galerkin Methods:
  Algorithms, Analysis, and Applications}.
\newblock New York: Springer-Verlag, 2008.

\bibitem{Huynh2007}
H.~T. Huynh, ``{A flux reconstruction approach to high-order schemes including
  discontinuous Galerkin methods},'' in {\em the 18th AIAA Computational Fluid
  Dynamics Conference}, (Miami, FL), 2007.
\newblock AIAA-2007-4079.

\bibitem{Huynh2009}
H.~T. Huynh, ``{A reconstruction approach to high-order schemes including
  discontinuous Galerkin methods for diffusion},'' in {\em the 47th AIAA
  Aerospace Sciences Meeting including The New Horizons Forum and Aerospace},
  (Orlando, FL), 2009.
\newblock AIAA-2009-403.

\bibitem{ZJWang2009}
{Z. J. Wang and H. Y. Gao}, ``{A unifying lifting collocation penalty
  formulation including the discontinuous Galerkin, spectral volume/difference
  methods for conservation laws on mixed grids},'' {\em Journal of
  Computational Physics}, vol.~228, pp.~8161--8186, 2009.

\bibitem{Vincent2011}
{P. E. Vincent, P. Castonguay and A. Jameson}, ``{A new class of high-order
  energy stable flux reconstruction schemes},'' {\em Journal of Scientific
  Computing}, vol.~47, pp.~50--72, 2011.

\bibitem{Bassi2005}
{F. Bassi, A. Crivellini, Stefano Rebay, and M. Savini}, ``{Discontinuous
  Galerkin solution of the Reynolds-averaged Navier–Stokes and $k–\omega$
  turbulence model equations},'' {\em {Computer \& Fluids}}, vol.~34,
  pp.~507--540, 2005.

\bibitem{Liang2009}
C.~Liang, S.~Premasuthana, A.~Jameson, and Z.~J. Wang, ``{Large Eddy Simulation
  of Compressible Turbulent Channel Flow with Spectral Difference method},'' in
  {\em 47th AIAA Aerospace Sciences Meeting including The New Horizons Forum
  and Aerospace Exposition}, 2009.
\newblock AIAA-2009-402.

\bibitem{Persson11}
{A. Uranga and P.‐O. Persson and M. Drela and J. Peraire}, ``{Implicit Large
  Eddy Simulation of transition to turbulence at low Reynolds numbers using a
  Discontinuous Galerkin method},'' {\em {International Journal for Numerical
  Methods in Engineering}}, vol.~87, pp.~232–--261, 2011.

\bibitem{bassi2016development}
F.~Bassi, L.~Botti, A.~Colombo, A.~Crivellini, A.~Ghidoni, and F.~Massa, ``{On
  the development of an implicit high-order Discontinuous Galerkin method for
  DNS and implicit LES of turbulent flows},'' {\em European Journal of
  Mechanics-B/Fluids}, vol.~55, pp.~367--379, 2016.

\bibitem{Fidkowski2016}
M.~A. Ceze and K.~J. Fidkowski, ``{High-Order Output-Based Adaptive Simulations
  of Turbulent Flow in Two Dimensions},'' {\em AIAA Journal}, vol.~54,
  pp.~2611--2625, 2016.

\bibitem{Wang17}
{Z.J. Wang and Y. Li and F. Jia and G. M. Laskowski and J. Kopriva and U.
  Paliath and R. Bhaskaran}, ``{Towards industrial large eddy simulation using
  the FR/CPR method},'' {\em {Computers \& Fluids}}, vol.~156, pp.~579--589,
  2017.

\bibitem{Vincent17}
{J. S. Park and F. D. Witherden and P. E. Vincent}, ``{High-Order Implicit
  Large-Eddy Simulations of Flow over a NACA0021 Aerofoil},'' {\em {AIAA
  Journal}}, vol.~55, pp.~2186--2197, 2017.

\bibitem{Marvriplis2018}
B.~R. Ahrabi, M.~J. Brazell, and D.~J. Mavriplis, ``{An Investigation of
  Continuous and Discontinuous Finite-Element Discretizations on Benchmark 3D
  Turbulent Flows},'' in {\em 2018 AIAA Aerospace Sciences Meeting}, 2018.
\newblock AIAA-2018-1569.

\bibitem{gottlieb2001strong}
S.~Gottlieb, C.-W. Shu, and E.~Tadmor, ``{Strong stability-preserving
  high-order time discretization methods},'' {\em SIAM review}, vol.~43, no.~1,
  pp.~89--112, 2001.

\bibitem{Butcher2002}
J.~C. Butcher, {\em Numerical Methods for Ordinary Differential Equations}.
\newblock Chichester: Wiley, 2002.

\bibitem{jameson1991time}
A.~Jameson, ``{Time dependent calculations using multigrid, with applications
  to unsteady flows past airfoils and wings},'' in {\em 10th Computational
  Fluid Dynamics Conference}, p.~1596.

\bibitem{jameson2017evaluation}
A.~Jameson, ``{Evaluation of Fully Implicit Runge Kutta Schemes for Unsteady
  Flow Calculations},'' {\em Journal of Scientific Computing}, vol.~73,
  no.~2-3, pp.~819--852, 2017.

\bibitem{Kennedy2016}
C.~A. Kennedy and M.~H. Carpenter, ``Diagonally implicit runge-kutta methods
  for ordinary differential equations. a review,'' Tech. Rep.
  NASA/TM–2016–219173, NASA.

\bibitem{vermeire2017behaviour}
B.~Vermeire and P.~Vincent, ``{On the behaviour of fully-discrete flux
  reconstruction schemes},'' {\em Computer Methods in Applied Mechanics and
  Engineering}, vol.~315, pp.~1053--1079, 2017.

\bibitem{bijl2002implicit}
H.~Bijl, M.~H. Carpenter, V.~N. Vatsa, and C.~A. Kennedy, ``{Implicit time
  integration schemes for the unsteady compressible Navier--Stokes equations:
  laminar flow},'' {\em Journal of Computational Physics}, vol.~179, no.~1,
  pp.~313--329, 2002.

\bibitem{wang2007implicit}
L.~Wang and D.~J. Mavriplis, ``{Implicit solution of the unsteady Euler
  equations for high-order accurate discontinuous Galerkin discretizations},''
  {\em Journal of Computational Physics}, vol.~225, no.~2, pp.~1994--2015,
  2007.

\bibitem{Baker1988}
A.~J. Baker and G.~S. Iannelli, ``{A stiffly-stable implicit Runge-Kutta
  algorithm for CFD applications},'' in {\em 26th AIAA Aerospace Sciences
  Meeting}, 1988.
\newblock AIAA Paper 88-0416.

\bibitem{lang2001ros3p}
J.~Lang and J.~Verwer, ``{ROS3P—an accurate third-order Rosenbrock solver
  designed for parabolic problems},'' {\em BIT Numerical Mathematics}, vol.~41,
  no.~4, pp.~731--738, 2001.

\bibitem{rang2005new}
J.~Rang and L.~Angermann, ``{New Rosenbrock W-methods of order 3 for partial
  differential algebraic equations of index 1},'' {\em BIT Numerical
  Mathematics}, vol.~45, no.~4, pp.~761--787, 2005.

\bibitem{rang2014analysis}
J.~Rang, ``{An analysis of the Prothero--Robinson example for constructing new
  DIRK and ROW methods},'' {\em Journal of Computational and Applied
  Mathematics}, vol.~262, pp.~105--114, 2014.

\bibitem{steinebach1995order}
G.~Steinebach, ``{Order-reduction of ROW-methods for DAEs and method of lines
  applications},'' 1995.

\bibitem{tranquilli2014rosenbrock}
P.~Tranquilli and A.~Sandu, ``{Rosenbrock--Krylov Methods for Large Systems of
  Differential Equations},'' {\em SIAM Journal on Scientific Computing},
  vol.~36, no.~3, pp.~A1313--A1338, 2014.

\bibitem{rang2015improved}
J.~Rang, ``{Improved traditional Rosenbrock--Wanner methods for stiff ODEs and
  DAEs},'' {\em Journal of Computational and Applied Mathematics}, vol.~286,
  pp.~128--144, 2015.

\bibitem{Bassi2015}
{F. Bassi, L. Botti, A. Colombo, A Ghidoni and F. Massa}, ``{Linearly implicit
  Rosenbrock-type Runge–Kutta schemes applied to the Discontinuous Galerkin
  solution of compressible and incompressible unsteady flows},'' {\em Computers
  \& Fluids}, vol.~118, pp.~305--320, 2015.

\bibitem{liu2016comparative}
X.~Liu, Y.~Xia, H.~Luo, and L.~Xuan, ``{A comparative study of Rosenbrock-type
  and implicit Runge-Kutta time integration for discontinuous Galerkin method
  for unsteady 3D compressible Navier--Stokes equations},'' {\em Communications
  in Computational Physics}, vol.~20, no.~4, pp.~1016--1044, 2016.

\bibitem{franciolini2017efficiency}
M.~Franciolini, A.~Crivellini, and A.~Nigro, ``{On the efficiency of a
  matrix-free linearly implicit time integration strategy for high-order
  Discontinuous Galerkin solutions of incompressible turbulent flows},'' {\em
  Computers \& Fluids}, vol.~159, pp.~276--294, 2017.

\bibitem{wang2018parallel}
L.~Wang and M.~Yu, ``{On the parallel implementation and performance study of
  high-order Rosenbrock-type implicit Runge-Kutta methods for the FR/CPR
  solutions of the Navier--Stokes equations},'' in {\em 2018 AIAA Aerospace
  Sciences Meeting}, 2018.
\newblock AIAA-2018-1095.

\bibitem{blom2016comparison}
D.~S. Blom, P.~Birken, H.~Bijl, F.~Kessels, A.~Meister, and A.~H. van Zuijlen,
  ``{A comparison of Rosenbrock and ESDIRK methods combined with iterative
  solvers for unsteady compressible flows},'' {\em Advances in Computational
  Mathematics}, vol.~42, no.~6, pp.~1401--1426, 2016.

\bibitem{sarshar2017numerical}
A.~Sarshar, P.~Tranquilli, B.~Pickering, A.~McCall, C.~J. Roy, and A.~Sandu,
  ``{A numerical investigation of matrix-free implicit time-stepping methods
  for large CFD simulations},'' {\em Computers \& Fluids}, vol.~159,
  pp.~53--63, 2017.

\bibitem{Yang2013}
H.~Yang, F.~Li, and J.~Qiu, ``{Dispersion and Dissipation Errors of Two Fully
  Discrete Discontinuous Galerkin Methods},'' {\em Journal of Scientific
  Computing}, vol.~55, pp.~552--574, 2013.

\bibitem{ZJW2018}
M.~Alhawwary and Z.~Wang, ``{Fourier analysis and evaluation of DG, FD and
  compact difference methods for conservation laws},'' {\em Journal of
  Computational Physics}, vol.~373, pp.~835--862, 2018.

\bibitem{Persson08}
{Persson, P.-O. and J. Peraire}, ``{Newton-GMRES Preconditioning for
  Discontinuous Galerkin Discretizations of the Navier--Stokes Equations},''
  {\em {SIAM Journal on Scientific Computing}}, vol.~30, pp.~2709--2733, 2008.

\bibitem{Roe81}
{P. L. Roe}, ``{Approximate Riemann solvers, parameter vectors and difference
  schemes},'' {\em {Journal of Computational Physics}}, vol.~43, pp.~357--372,
  1981.

\bibitem{gao2013differential}
H.~Gao, Z.~Wang, and H.~Huynh, ``{Differential formulation of discontinuous
  Galerkin and related methods for the Navier--Stokes equations},'' {\em
  Communications in Computational Physics}, vol.~13, no.~4, pp.~1013--1044,
  2013.

\bibitem{mulder1985experiments}
W.~A. Mulder and B.~Van~Leer, ``{Experiments with implicit upwind methods for
  the Euler equations},'' {\em Journal of Computational Physics}, vol.~59,
  no.~2, pp.~232--246, 1985.

\bibitem{knoll2004jacobian}
D.~A. Knoll and D.~E. Keyes, ``{Jacobian-free Newton--Krylov methods: a survey
  of approaches and applications},'' {\em Journal of Computational Physics},
  vol.~193, no.~2, pp.~357--397, 2004.

\bibitem{sharov2000implementation}
D.~Sharov, H.~Luo, J.~Baum, and R.~L{\"o}hner, ``{Implementation of
  unstructured grid GMRES+LU-SGS method on shared-memory, cache-based parallel
  computers},'' in {\em 38th Aerospace Sciences Meeting and Exhibit}, p.~927,
  2000.

\bibitem{franciolini2018p}
M.~Franciolini, L.~Botti, A.~Colombo, and A.~Crivellini, ``{$p$-Multigrid
  matrix-free discontinuous Galerkin solution strategies for the under-resolved
  simulation of incompressible turbulent flows},'' {\em arXiv preprint
  arXiv:1809.00866}, 2018.

\bibitem{van2011comparison}
W.~M. Van~Rees, A.~Leonard, D.~Pullin, and P.~Koumoutsakos, ``A comparison of
  vortex and pseudo-spectral methods for the simulation of periodic vortical
  flows at high reynolds numbers,'' {\em Journal of Computational Physics},
  vol.~230, no.~8, pp.~2794--2805, 2011.

\end{thebibliography}
\end{document}